\documentclass{elsart}

\usepackage{harvard}

\usepackage{graphicx}

\usepackage{amssymb}

\def\astrobj#1{}
\def\url#1{{\ttfamily\def\/{/\discretionary{}{}{}}#1}}

\begin{document}

\begin{frontmatter}

\title{A\,0620$-$00 revisited: a black-hole transient
case-study\thanksref{label1}}
\thanks[label1]{This paper is dedicated to one of the 
discoverers of the optical counterpart of A\,0620$-$00 (V616\,Mon, 
Nova Mon 1975), Dr.~Forrest Boley, who died on September 9, 1997.}

\author{Erik~Kuulkers}

\address{Astrophysics, University of Oxford, Nuclear and Astrophysics
Laboratory, Keble Road, Oxford OX1 3RH, United Kingdom\thanksref{email}}
\thanks[email]{E-mail: e.kuulkers1@physics.oxford.ac.uk}


\begin{abstract}
For the first time we have performed a detailed study of the X-ray, optical and
infra-red light curves and an intercomparison between these wavelengths 
(including radio) of the 1975/1976 outburst of the famous black-hole transient 
A\,0620$-$00 (Nova Mon 1975, V616\,Mon). We also investigated the optical
behaviour up to a year after the main outburst. 
This study enabled us to find some new features, which have not been 
discussed before.

During the various stages of the outburst of A\,0620$-$00
we found the X-rays lag the optical
on the order of $\sim$5 to $\sim$20~days. Moreover, we found evidence 
that the activity associated with the secondary maximum started even earlier
in the infra-red. This suggests that most of the processes associated with
the outburst occur in the outer parts of the accretion disk.

Although various drops in intensity (lasting on the order of a day or more) 
in the optical and X-ray outburst light curves of A\,0620$-$00 
have been reported before, we identified some new ones. One such X-ray `dip' 
only appeared in the soft X-rays
(1.5--6\,keV) whereas at higher energies ($\gtrsim$6\,keV) the intensity
slightly
increases. This shows that the X-ray spectrum pivots near $\sim$6\,keV around
that time. In the optical we found evidence for another local maximum 
around that time (so-called `intermediate' maximum). The `intermediate' maximum
appears $\sim$30~days after the secondary maximum, which is close to the
viscous time scale of an irradiated accretion disk. We suggest this feature
to be an `echo' of the secondary maximum.
At the end of the outburst another local maximum occurs. Since the time
difference between the peak of the tertiary maximum and this local maximum is
also on the order of the viscous time scale we suggest that this feature is 
an `echo' of the tertiary maximum.
We find that drops in the optical and X-ray intensity before and during the 
rise to tertiary maximum are also present in various other short period soft
X-ray transients (SXTs). They always occur $\sim$150~days after the start of the
outburst. 

Although the X-ray spectrum of A\,0620$-$00 gradually softens during the decay 
from outburst, we find for the first time that it starts to harden again 
$\sim$100--150 days after the start of the outburst, similar to that seen
in GS\,2000+25 and GS\,1124$-$68. This suggests we witness the transition
from the so-called high to low state identified in black-hole X-ray binaries.
We note that the time of spectral hardening in A\,0620$-$00, GS\,2000+25 and
GS\,1124$-$68 is simultaneous with the occurrence of the drops in optical and 
X-ray intensity.

We also show that the optical outburst amplitude and the shape of the optical 
outburst light curve of A\,0620$-$00 closely resembles those of the 
cataclysmic variable AL\,Com (where the compact star is a white dwarf). This 
strengthens the similarity in outburst and quiescent properties of the SXTs and 
`Tremendous Outburst Amplitude Dwarf novae' (TOADs) or WZ\,Sge stars, and 
shows that in general the optical outburst light curves of both groups are 
governed by the disk properties and not by the compact object.
Since irradiation provides a natural mechanism to prolong the outburst of SXTs,
we suggest this could be of influence as well during TOAD outbursts.

\end{abstract}

\end{frontmatter}



\section{Introduction}\label{intro}

In 1975 a powerful X-ray transient, \astrobj{A0620-00}{A\,0620$-$00}, emerged 
in the constellation 
Monoceros (Elvis et al.\ 1975). \astrobj{A0620-00}{A\,0620$-$00} was 
the first X-ray 
transient to be identified
with an optical brightening of a star, \astrobj{V616Mon}{V616\,Mon}, at the 
same position (Boley \&\ Wolfson 1975; Boley et al.\ 1976). It 
was therefore also designated as Nova Mon 1975. 
It is still the brightest extra-solar X-ray source seen to date, and 
many of the exciting results during the first months of 
its outburst were summarized by Maran (1976).

Because of its brightness the source has been studied in considerable detail at
various wavelengths. The results have appeared in a large numbers of papers 
in the literature. There exists, however, no thorough comparison of the 
optical (UBV) light curves together with the X-ray light curves from various
experiments. Either authors have presented (part of) the optical light curve 
(with sometimes the inclusion of data points from other observers) in one or 
two optical passbands and sometimes compared it with the X-ray light curve
(e.g.\ Robertson, Warren \&\ Bywater 1976; Shugarov 1976; 
Lloyd, Noble \&\ Penston 1977; Ciatti, Mammano \&\ Vittone 1977;
Tsunemi, Matsuoka \&\ Takagishi 1977; Chen, Livio \&\ Gehrels 1993; 
Goranskii et al.\ 1996).
The first rough comparison of optical and X-ray light curves 
was undertaken by Whelan et al.\ (1977), together with a crude comparison with 
radio, infra-red and ultra-violet measurements.

Fortunately, almost all the optical (plus infra-red and ultra-violet) observations have 
been compiled by Ron Webbink in his unpublished report in 1978. 
In this report he showed (only) the overall B band light curve, and the
corresponding B$-$V, and U$-$B measurements, whenever they were simultaneous.
Subsequently, only a few authors have shown his B band light curve
(e.g., Chevalier \&\ Ilovaisky 1990; 
Van Paradijs \&\ McClintock 1995).
Since the outburst of \astrobj{A0620-00}{A\,0620$-$00} more (soft) X-ray 
transients (SXTs) have been identified,
some of them with considerable optical coverage 
(see Chen, Shrader \&\ Livio 1997). We note that most of these
optical observations were only in one or two passbands, and so far 
\astrobj{A0620-00}{A\,0620$-$00} is still the only transient with a fairly 
complete optical light curve in more than 2 passbands (largely due to its 
brightness).
Several of the X-ray transients have been either found to contain a 
black hole, including \astrobj{A0620-00}{A\,0620$-$00} 
(McClintock \&\ Remillard 1986),
or a neutron star (e.g.\ Van Paradijs 1998). 
Moreover, the comparison of the various outburst 
light curves (see White, Kaluzienski \&\ Swank 1984;
Van Paradijs \&\ Verbunt 1984; Chen et al.\ 1993);
Van Paradijs \&\ McClintock 1995;
Tanaka \&\ Lewin 1995; Tanaka \&\ Shibazaki 1996; 
Chen et al.\ 1997), and the short term ($\lesssim$100\,s) (X-ray) variability
(e.g.\ Tanaka \&\ Lewin 1995; van der Klis 1995) of the 
X-ray transients has led to a general picture of their behaviour,
in which the main driver for these characteristics is the mass accretion
rate through the accretion disk.
Since no thorough study of the outburst light curve of 
\astrobj{A0620-00}{A\,0620$-$00} has been performed to date,
and the fact that we can now place the results of its 1975/1976 outburst 
within the general 
framework of X-ray transient behaviour, we decided to perform, for the first 
time, an analysis of all the available optical, infra-red and X-ray light 
curves of \astrobj{A0620-00}{A\,0620$-$00} and 
undertake a detailed comparison between them.

In the next Sections we will first give an overview of what is already known of 
\astrobj{A0620-00}{A\,0620$-$00}, then present the X-ray, optical and
infra-red outburst light curves with newly 
identified features, and then compare it with behaviour seen in other X-ray 
transients. Finally, we will compare the properties of 
\astrobj{A0620-00}{A\,0620$-$00}
with a special unstable class of the cataclysmic variables, i.e., 
the `Tremendous Outburst Amplitude Dwarf novae' (TOADs), or the 
\astrobj{WZSge}{WZ\,Sge} stars.

\section{A\,0620$-$00 revisited}\label{a0620_intro}

\subsection{Outburst}

\astrobj{A0620-00}{A\,0620$-$00} was discovered by the Sky Survey Experiment 
on board \emph{Ariel~V} on August 3, 1975 (Elvis et al.\ 1975).
Subsequently, after a more precise location 
became available from \emph{SAS-3} (Doxsey et al.\ 1976), the source 
was detected in the radio (e.g.\ Owen et al.\ 1976), optical 
(Boley et al.\ 1976) and infra-red 
(e.g.\ Kleinmann, Brecher \&\ Ingham 1976). 
The reason why \astrobj{A0620-00}{A\,0620$-$00} was so bright during outburst 
is its proximity, with an estimated distance of $\sim$1\,kpc 
(see e.g.\ Shahbaz, Naylor \&\ Charles 1994).

During the rise to X-ray maximum a precursor was seen.
The maximum of the X-ray outburst was reached within a week after the 
precursor (Elvis et al.\ 1975). The subsequent decay was approximately
exponential with a decay time scale of $\sim$29~days up to the beginning of
October (Kaluzienski et al.\ 1977; Pounds et al.\ 1977). About 
one and a half months after X-ray maximum the decay halted for $\sim$10~days
during which the intensity slightly increased (Carpenter et al.\ 
1976; Matilsky et al.\ 1976; Kaluzienski et al.\ 1977); 
this period is called a `glitch' or secondary maximum in the decay light curve
(see e.g.\ Chen et al.\ 1997). Thereafter the exponential decay resumed again 
with a time scale of $\sim$20~days (Kaluzienski et al.\ 1977; 
Pounds et al.\ 1977), until the beginning of 1976 when the outburst light
curve flattened. At the end of 1976 January
the X-ray flux increased again for about a month, and subsequently 
faded rapidly with a decay time scale of less than $\sim$10~days to below 
detection limits (Kaluzienski et al.\ 1976; 1977). This 
last episode showed up as a large `bump' or a broad (tertiary) maximum at the 
end of the outburst light curve.

In the radio \astrobj{A0620-00}{A\,0620$-$00} was only active for a few weeks 
(e.g.\ Davis et al.\ 1975; see also Hjellming et al.\ 1988).
The radio decay time was $\sim$5~days (Davis et al.\ 1975).

In the optical, \astrobj{A0620-00}{A\,0620$-$00} faded from maximum less fast 
than in X-rays by about 0.015 magnitude per day, corresponding to a 
decay time scale of $\sim$67~days (Duerbeck \&\ Walter 1976; 
Whelan et al.\ 1977; Lloyd et al.\ 1977; 
Pounds et al.\ 1977; see also Tsunemi et al.\ 1977; 
Goranskii et al.\ 1996). 
The glitch also showed up in the optical, at approximately the 
same time as in X-rays (e.g.\ Kaluzienski et al.\ 1977; 
Whelan et al.\ 1977; Tsunemi et al.\ 1977; see also Chen et al.\ 1993). 
A large bump could also be seen at the end of the optical outburst light 
curve (e.g.\ Lloyd et al.\ 1977; Whelan et al.\ 1977; Tsunemi et al.\ 1977), 
but this time the optical rose before the X-ray did
(e.g.\ Chen et al.\ 1993). 
The final optical decay started about $\sim$2~weeks later than the final 
X-ray decay, and had a fast decay time scale of $\sim$12~days
(Tsunemi et al.\ 1977). UBV measurements during the final decay 
showed B$-$V already near its quiescent value, whereas U$-$B was still at 
its outburst value (Lyutyi 1976; see also 
Lyutyi \&\ Shugarov 1979). 

A $\sim$3.9~day modulation in the optical UBV was reported during the first 
part of the decay up to the glitch (Duerbeck \&\ Walter 1976). 
However, a modulation on a time-scale
of $\sim$8--10~days just after the first glitch, was reported
by Chevalier et al.\ (1980). Moreover, Tsunemi et al. (1977)
reported an optical period of $\sim$7.4 days from the three months
during the decay. Subsequent reanalysis of the data of Duerbeck \&\ Walter
(1976) did not reproduce the $\sim$3.9~day period (Matilsky et al.\
1976; Chevalier et al.\ 1980; see also 
Tsunemi et al.\ 1977). Minima in the optical light 
curve below the extrapolated general decay (Lloyd et al.\ 1977, 
Robertson et al.\ 1976) and an eclipse-like feature near the end of the 
outburst (Chevalier et al.\ 1980) were consistent with the
$\sim$7.4~day variability. 

In X-rays a period of $\sim$7.5~days during the main decay and $\sim$7.8~days
during the `trough' before the broad tertiary maximum plus initial 
rise to the tertiary maximum was found in \emph{SAS-3} data 
(Matilsky et al.\ 1976).
A reanalysis of the \emph{SAS-3} data, however, did not reveal the variability 
in the earlier phase of the decline, but rather a smooth exponential 
decay; the $\sim$7.8~days variability was confirmed (Plaks 1991).
Intensity minima found in the overall light curve of the 
various \emph{Ariel V} instruments (Pounds et al.\ 1977;
see Watson 1982) are related to the 7.8-day cycle.
The amplitude of the modulation is largest when 
\astrobj{A0620-00}{A\,0620$-$00} is weakest
(Matilsky et al.\ 1976; Pounds et al.\ 1977).
It was recognized by Tsunemi et al.\ (1977) that the optical 7.8-day 
modulation was not in phase with that in X-rays; 
the X-rays were reported to precede the optical by about one day 
(Chevalier et al.\ 1980).

During the outburst the optical revealed B$-$V and U$-$B colours
very similar to \astrobj{ScoX-1}{Sco\,X-1} and \astrobj{CygX-2}{Cyg\,X-2} 
(e.g.\ Eachus, Wright \&\ Liller 1976; 
Matsuoka \&\ Tsunemi 1976).
Also the optical spectra at outburst maximum were similar to 
\astrobj{ScoX-1}{Sco\,X-1}, showing
evidence for X-ray heating effects. However, the emission features of 
H$\alpha$ and H$\beta$ were more like dwarf novae in outburst 
(Whelan et al.\ 1977).
During the maximum and initial decline phase the ratio of the 
X-ray to optical luminosity was $\sim$1000, again showing the similarity
to \astrobj{ScoX-1}{Sco\,X-1} 
(Boley et al.\ 1976; Whelan et al.\ 1977; Matilsky et al.\ 1976; 
Kaluzienski et al.\ 1977; see also Tsunemi et al.\ 1977).

The X-ray spectral behaviour of \astrobj{A0620-00}{A\,0620$-$00} was that of 
a typical X-ray transient (see e.g.\ Tanaka \&\ Lewin 1995; 
Tanaka \&\ Shibazaki 1996). Before the precursor the spectrum was hard, 
whereas during the subsequent rise to maximum the spectrum rapidly softened 
(Ricketts, Pounds \&\ Turner 1975; Matilsky et al.\ 1976); 
the X-ray flux above $\sim$10\,keV actually fell during that period 
(Ricketts et al.\ 1975).
Near X-ray maximum there was evidence for a hard component above $\sim$15\,keV
(Ricketts et al.\ 1975; see also White et al.\ 1984), and possibly
even above $\sim$30\,keV (Coe, Engel \&\ Quenby 1976). 
During the main decay the spectral hardness gradually declined
(Matilsky et al.\ 1976, Carpenter et al.\ 1976). 
Two months after X-ray maximum Coe et al.\ (1976) reported spectral hardening 
(up to $\sim$200\,keV) with respect to their measurement near X-ray maximum.
However, as already noted by e.g.\ Tanaka \&\ Lewin (1995), their 
measurements are very marginal. In fact, a study of the high energy range
up to $\sim$56\,keV, as measured with \emph{SAS-3} nearly simultaneously with 
the measurements by Coe et al.\ (1976), did not reveal evidence for
such a hard power-law tail (Hwang 1988).

On short timescales (from hours down to msec) 
\astrobj{A0620-00}{A\,0620$-$00} was very quiet during its outburst.
No modulation larger than $\sim$3\%\/ over timescales 200\,s--2\,d 
was seen during the rise and at maximum (\emph{Ariel~V} SSE, 2--18\,keV, 
Elvis et al.\ 1975).
Near X-ray maximum also down to 0.2\,msec no evidence for pulsations 
were reported (\emph{SAS-3} slat collimator system, 1.5--60\,keV, 
Doxsey et al.\ 1976).
During the first months after maximum there was no evidence for periodic 
fluctuations or the shot noise variability seen in e.g.\ 
\astrobj{ScoX-1}{Sco\,X-1} 
(\emph{Ariel~V} RMC, 3--8\,keV, Carpenter et al.\ 1976).
Measurements by the medium X-ray energy detector (1--8\,keV) on board the 
{\it Astronomical Netherlands Satellite} (ANS)
during the glitch also revealed no indications of periodic variations between 
0.25 and 100\,s (Brinkman et al.\ 1976).
The high count rates observed with the various X-ray detectors 
from \astrobj{A0620-00}{A\,0620$-$00} would have definitely revealed any 
variability similar 
to that seen in the low state or very high state of black-hole candidates 
(e.g.\ van der Klis 1995, and references therein). 
Since none were reported and from the fact that the X-ray spectrum was very 
soft, White et al.\ (1984) already concluded that
\astrobj{A0620-00}{A\,0620$-$00} was most probably in its high state during 
most of the
observations. We note that this is consistent with the behaviour 
of \astrobj{A0620-00}{A\,0620$-$00} with respect to other SXTs in the hard 
(20--200\,keV) versus 
soft (1--20\,keV) X-ray luminosity study of Barret, 
McClintock \&\ Grindlay (1996).
The spectral changes during the rise of the outburst 
may have indicated a change from the low to a high state 
(White et al.\ 1984).

Optical observations just after the glitch on 1975 October 10 
showed no random variations with amplitudes larger 
than 0.01 magnitudes between 2 and 200 sec (Robinson \&\ Nather 1975).
Also during several nights in March 1976 (i.e.\ during the large bump at the 
end of the outburst) no optical flickering 
activity was found down to $\sim$2\,s (Chevalier et al.\ 1980). 

\subsection{Quiescence}

Soon after the first optical detection of \astrobj{A0620-00}{A\,0620$-$00}, 
the quiescent
optical star (\astrobj{V616Mon}{V616\,Mon}) was found on archival photographic 
plates at a magnitude of B$_{\rm pg}\sim 20.0$--$20.5$ 
(Ward et al.\ 1975;
Eachus et al.\ 1976). Together with the magnitude near
optical maximum this led to a large outburst amplitude of $\sim$9~mag.
Even earlier photographic plates revealed the existence of a previous
outburst of \astrobj{A0620-00}{A\,0620$-$00} at the end of 1917 
(Eachus et al.\ 1976),
suggesting an outburst recurrence time of $\sim$60~years.

\astrobj{A0620-00}{A\,0620$-$00} was quickly recognized to have a binary 
nature, most probably 
related to dwarf novae.
Based on its transient nature and its brightness in X-rays it was suggested 
that \astrobj{A0620-00}{A\,0620$-$00} contained a neutron star or a black hole 
instead of a white dwarf (e.g.\ Elvis et al.\ 1975; 
Avni, Fabian \&\ Pringle 1976).
From photometric colours in quiescence (using plates taken before the 
1975/1976 outburst) Ward et al.\ (1975) suggested the companion star 
to be a red dwarf, most likely a K star. 
Subsequently, it was spectroscopically
identified as a K5V star (Oke 1977; McClintock et al.\ 1983; 
see also Murdin et al.\ 1980).

It was suggested that if the companion star was a red dwarf its orbital
period would be $\sim$8\,hr (Avni et al.\ 1976).
When \astrobj{A0620-00}{A\,0620$-$00} had reached quiescence, the search 
for its orbital period 
was started. It was finally established to be $\sim$7.75\,hr 
(McClintock et al.\ 1983; McClintock \&\ Remillard 1986).

The first dynamical evidence for the compact star being a black black hole
came from the work of McClintock \&\ Remillard (1986). They found
a compact star mass of M$_{\rm X}>3.2$\,M$_{\odot}$, with a most likely 
value of $\sim$7\,M$_{\odot}$.
By measuring the rotational broadening of the absorption line spectrum 
the mass ratio\footnote{Mass ratio $q$$\equiv$M$_{\rm opt}$/M$_{\rm X}$, 
where M$_{\rm opt}$ and M$_{\rm X}$ are the 
mass of the companion star and compact object, respectively.},
was found to be $\sim$0.07 (Marsh, Robinson \&\ Wood 1994). 
Finally, the analysis of ellipsoidal variations in the infra-red light 
curves revealed the inclination to be $\sim$37$^{\circ}$, which therefore
implied the mass of the compact star and the companion star to be
M$_{\rm X}\sim 10$\,M$_{\odot}$ and M$_{\rm opt}\sim 0.6$\,M$_{\odot}$,
respectively (Shahbaz et al.\ 1994).
We note that the quiescent BV light curves are somewhat more complex
(see also Sect.~\ref{TOADs}), and that their irregularities have been
interpreted as a grazing eclipse of the companion star by the accretion disk;
this would imply inclinations of $i>62^{\circ}$ (Haswell et al.\ 1993).

\section{Observations}

\subsection{X-ray}

The X-ray outburst light curves shown in this paper are those from 
instruments on board {\it SAS-3} and {\it Ariel~V}. 
Data from the {\it SAS-3} Center Slat collimator (CSL) provided
information in four energy channels (CSL A--D) from 1.5--60\,keV
(Buff et al.\ 1977). Parts of these data have been
reported by Doxsey et al.\ (1976), Matilsky et al.\ (1976)
and Plaks (1991). We used the data as re-analysed by Plaks (1991), 
i.e.\ data from the two argon channels, 1.5--6\,keV and 6--15\,keV, and 
data from the first xenon channel, 10--42\,keV.

The {\it Ariel~V} instruments from which we have used data are the 
Sky Survey Experiment (SSE, also named as Sky Survey Instrument or SSI), the 
All Sky Monitor (ASM), and the Rotation Modulation Collimator (RMC).
The SSE provided count rates in eight channels covering 2--18 keV 
(Villa et al.\ 1976). We used the lightcurve 
from all eight channels combined as presented by Elvis et al.\ 
(1975; see also Chen et al.\ 1997) 
and Watson (1982).
The ASM data are those presented by
Kaluzienski et al.\ (1977; see also Chen et al.\ 1997) 
and give the count rate in one 3--6\,keV energy band (Holt 1976). 
The RMC provided data in three energy channels, i.e.\ 3.0--4.3\,keV, 
2.3--5.9\,keV and 4.9--7.6\,keV. We used the data in the combined 3--7.6\,keV 
energy band, as presented by Carpenter et al.\ (1976). 
A composite outburst light curve of all the 
{\it Ariel~V} SSE, ASM, and RMC data, together with the {\it SAS-3} coverage 
was given by Watson (1982).
We note that more sketchy overall X-ray outburst light curves have been given 
by Carpenter et al.\ (1976) and White et al.\ (1984).

\subsection{Optical}\label{optical_observations}

Almost all of the optical data presented in this paper come from the
compilation by Webbink (1978). They include all photoelectric, 
photographic,
and visual observations appearing in the
published literature available to him, and many unpublished observations 
communicated to him, up to the beginning of 1977. Observations noted as 
uncertain, doubtful, or erroneous (see Webbink 1978) have not been taken into 
account in this paper. We also did not use the data by Tsunemi et al.\ 
(1977), since their measurements are known to have systematic 
differences with respect to the other B band data (see 
Tsunemi et al.\ 1977). 
From our analysis we find that this also applies to the
photographic B measurements of Matsuoka \&\ Tsunemi (1976).
In addition we used the data obtained during outburst as reported 
by Hudec (B$_{\rm pg}$, 1977), 
Borisov, Derevyashkin \&\ Deych (B$_{\rm pg}$, 1977) and 
Chevalier et al.\ (UBV, 1980), and the data obtained in 
quiescence up to $\sim$1~year after the final decay from the outburst 
as reported by Lyutyi \&\ Shugarov (V$_{\rm pg}$, 1979) and Kurochkin, 
Karitskaya \&\ Bochkarev (B$_{\rm pg}$, 1988).

Apart from the photographic B and V plates obtained with the 26- and 13-inch 
reflectors of the Royal Greenwich Observatory during the period 1975 August 27 
and 1976 April 4 at Herstmonceux (Lloyd et al.\ 1977), there were 2 
additional photographic B plates which have not been reported before.
They were taken with the 26-inch reflector on 1976 October 21 (03:29 UT) and 
October 22 (02:56 UT). At both times the optical star was 
not visible, resulting in upper limits of B$_{\rm pg}\sim 16.5$ and 
B$_{\rm pg}\sim 15.0$, respectively (Argyle 1997, private communication).

The visual measurements of observers of the Variable Star Observers League in
Japan (VSOLJ) were not available in Webbink's (1978) compilation, and we 
have therefore included them in our analysis. We also used additional  
visual measurements (mainly upper limits) after the final decline from 
outburst by observers from the American Association of Variable Star Observers 
(AAVSO, Mattei 1997), the Variable Star Section of the Royal Astronomical 
Sociey of New Zealand (VSS RAS NZ), and the Association Francaise des
Observateurs d'Etoiles Variables (AFOEV).

The uncertainties in the photoelectric UBV observations during outburst 
were typically $\sim$0.01--0.03, mainly depending on the 
brightness of \astrobj{A0620-00}{A\,0620$-$00} and the observed wavelength band 
(see Duerbeck \&\ Walter 1976, Robertson et al.\ 1976; 
Lyutyi 1976; Chevalier et al.\ 1980). During the final decline to quiescence 
this uncertainty increased to $\sim$0.05--0.1, depending on the wavelength band 
(Lyutyi 1976).
The uncertainties in the photographic BV observations during outburst were
typically $\sim$0.07--0.1 (e.g.\ Lloyd et al.\ 1977), and sometimes
$\sim$0.1--0.2 (Hudec 1977), depending on the instrument used.
In quiescence the typical uncertainties were $\sim$0.1--0.4 for the
photoelectric UBV observations, depending on the instrument used and the
wavelength band (see Ciatti et al.\ 1977; Lyutyi \&\ Shugarov 1979),
and $\sim$0.2--0.5 for the photographic BV observations
(Lyutyi \&\ Shugarov 1979; Kurochkin et al.\ 1988).
The visual estimates had usual uncertainties of $\sim$0.1 during outburst
for one observer. However, the systematic uncertainties between different 
observers are larger mainly because of the differences in the 
observers themselves, the inadequacy 
of finding charts used at that time and the systematic errors in the magnitudes
of comparison stars. 

\section{Results}\label{results}

\subsection{Outburst}\label{outburst}

\subsubsection{X-ray behaviour}\label{x-ray}

\begin{figure}
\centerline{
\includegraphics*[bb = 58 109 534 588,width=12cm]{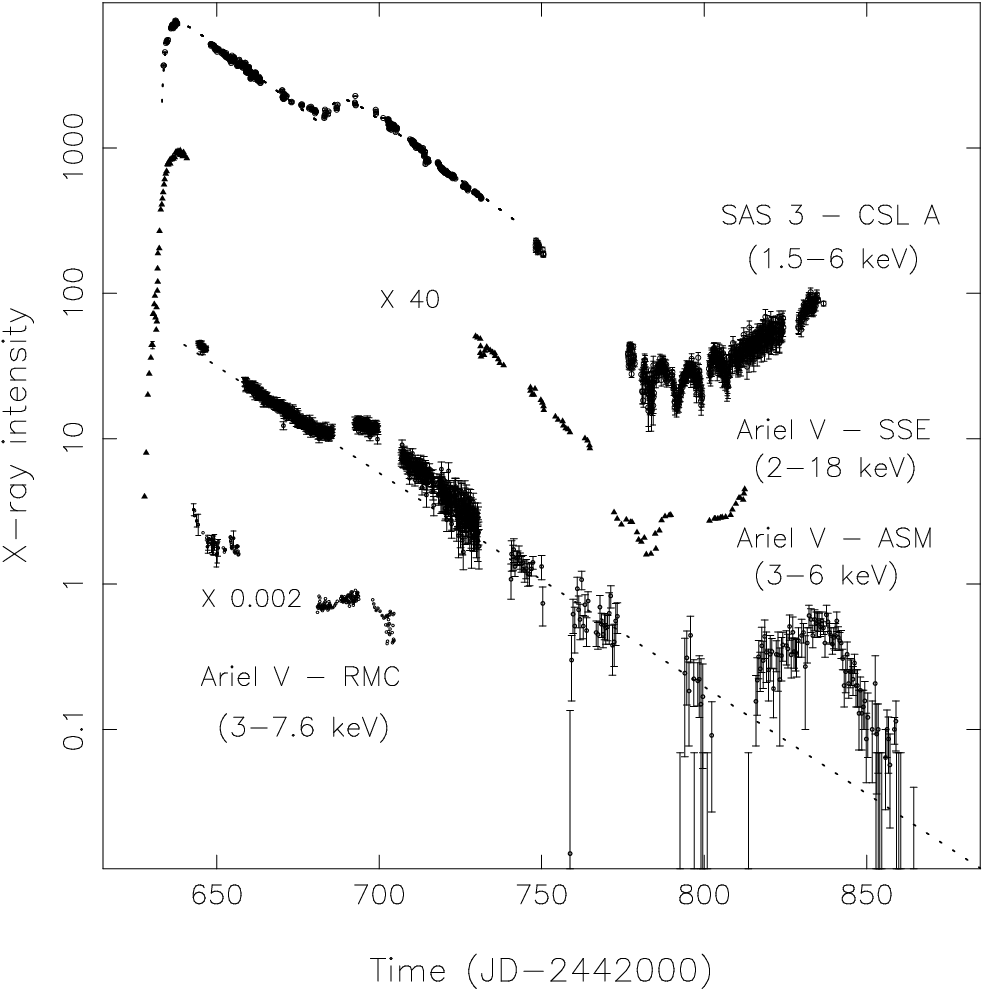}}
\caption*{\small X-ray outburst light curves of \astrobj{A0620-00}{A\,0620$-$00} as 
obtained with (from top to bottom) the
{\it SAS-3} CSL-A (1.5--6\,keV), and the {\it Ariel~V} SSE (2--18\,keV), ASM
(3--6\,keV) and RMC (3--7.6\,keV) instruments. The units of the {\it SAS-3}
CSL-A and the {\it Ariel~V} RMC data are cts\,s$^{-1}$, whereas that of
the {\it Ariel~V} SSE and ASM are given in units of the Crab count rate
(in the corresponding energy bands).
Note that the {\it Ariel~V} SSE has been shifted by a factor of 40 and the 
RMC data by a factor of 0.002. Error bars on each measurement are given for all
data of the {\it SAS-3} CSL-A and {\it Ariel~V} ASM instruments, whereas 
for the {\it Ariel~V} SSE typical error bars are given at the beginning of the 
outburst and near the maximum (see Elvis et al.\ 1975); when no errors
are plotted for the {\it Ariel~V} RMC data, they are less than 10\%\ (see 
Carpenter et al.\ 1976). For the first $\sim$110~days of the 
{\it SAS-3} CSL-A light curve we show the corresponding fit to the 
model of Augusteijn, Kuulkers \&\ Shaham (1993, see text for
parameters), while for the {\it Ariel~V} ASM we show the fit to part of the 
first exponential decline (JD\,2442658--JD\,2442682) and its extrapolation to 
the beginning and the end of the outburst.}
\label{a0620_allxray}
\end{figure}

\begin{figure}
\centerline{
\includegraphics*[bb = 58 109 534 701,width=11cm]{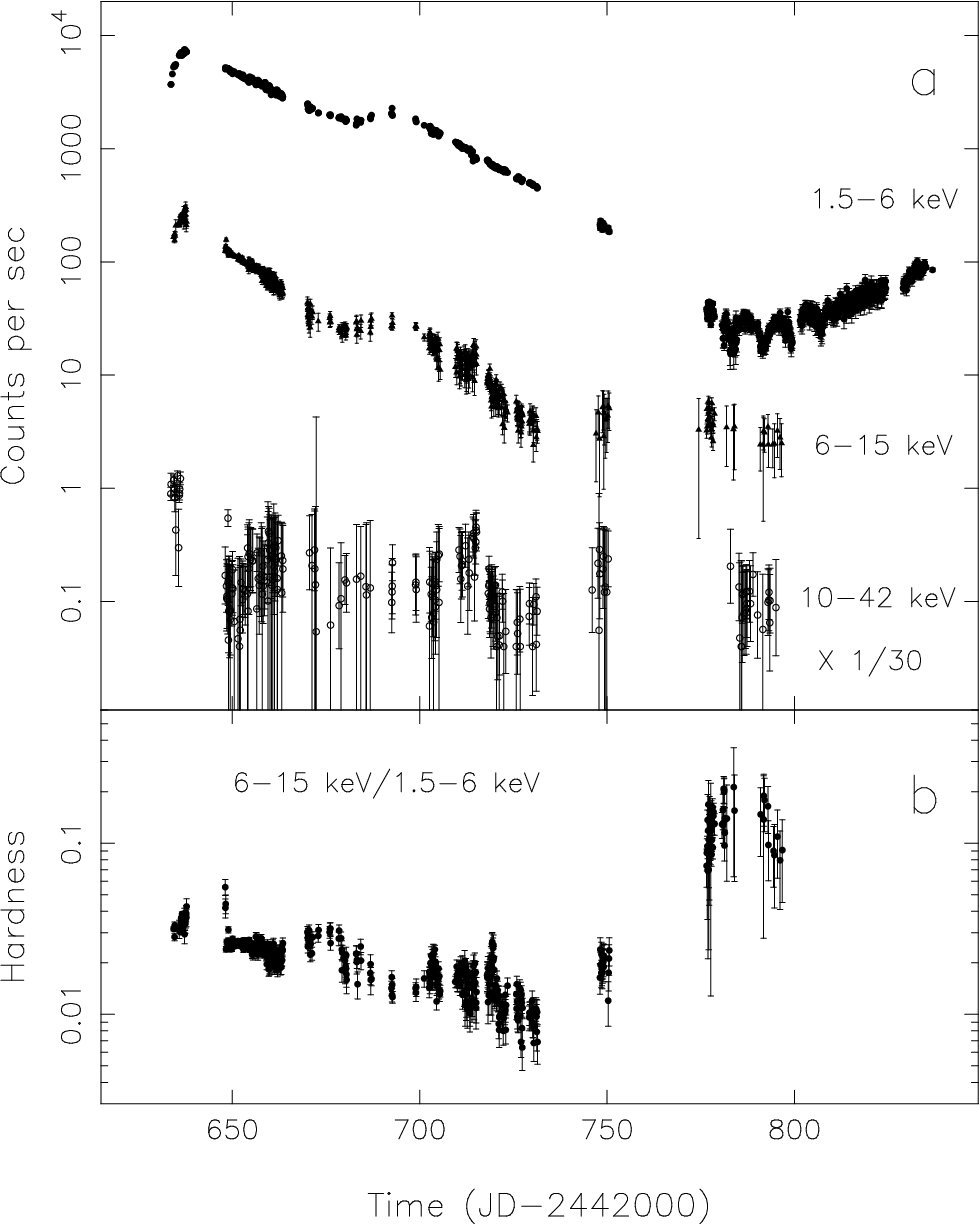}}
\caption{\small \textbf{a.} X-ray outburst light curves of 
\astrobj{A0620-00}{A\,0620$-$00} as obtained with (from top to bottom) 
the {\it SAS-3} CSL-A (1.5--6\,keV), CSL-B (6--15\,keV) and CSL-C (10--42\,keV)
detectors. The CSL-C light curve has been offset by a factor of 1/30 for
clarity. \textbf{b.} Hardness light curve of \astrobj{A0620-00}{A\,0620$-$00}. 
Hardness is defined as
the ratio of the count rates from the CSL-B and the CSL-A detector. Note the
clear hardening after JD\,24462732.}
\label{a0620_sas3}
\end{figure}

In Fig.~\ref{a0620_allxray} we show the X-ray outburst light curves of 
of \astrobj{A0620-00}{A\,0620$-$00} obtained with four instruments, i.e., 
the {\it SAS-3} CSL-A (1.5--6\,keV), and the {\it Ariel~V}
SSE (2--18\,keV), ASM (3--6\,keV) and RMC (3--7.6\,keV). In 
Fig.~\ref{a0620_sas3}a
we show the {\it SAS-3} CSL-A data again, together with CSL-B (6--15\,keV) and 
CSL-C (10--42\,keV) data.
Clearly, the light curves show the fast rise to maximum,
the exponential decay for $\sim$45~days, the secondary 
maximum $\sim$60~days after the start of the outburst followed by a faster
exponential decay with respect to the first part, the $\sim$7.8~day 
oscillations during the trough, the broad tertiary maximum $\sim$200~days
after start of the outburst and the final fast decay to below the detection
limits. 

The maximum of the precursor in the {\it Ariel~V} SSE light curve was 
at JD\,2442631 (Elvis et al.\ 1975; see also 
Fig.~\ref{a0620_detail}a). We note that this feature could 
also represent a drop in intensity 
(similar to that seen later in the outburst, see 
below), with a minimum at JD\,2442631.4 and a duration of about a day. 
The maximum of the outburst was reached near JD\,2442637 
(see Elvis et al.\ 1975; Doxsey et al.\ 1976), whereas the secondary 
maximum peaked near JD\,2442692 (see Matilsky et al.\ 1976, see also
Fig.~\ref{a0620_detail}b). During the 
first and second exponential decays the source slowly softened as can be 
seen in Fig.~\ref{a0620_sas3}b where we
plot the hardness ratio using the {\it SAS-3} CSL-A and CSL-B detectors
(see also Matilsky et al.\ 1976; Plaks 1991).

In Fig.~\ref{a0620_allxray} we also show the result of an exponential fit 
to part of the first exponential decline (JD\,2442658--JD\,2442682) of 
the {\it Ariel V} ASM light curve and its extrapolation to the beginning 
and the end of the outburst. A qualitatively similar fit is obtained 
from the same stage of the {\it SAS-3} CSL-A light curve 
(see Fig.~\ref{a0620_detail}b). As can 
be seen, the secondary and tertiary maxima are enhancements in the
lightcurve; the X-ray intensity seems to always return to the extrapolated 
level expected from the first exponential decay. The drops in intensity 
related to the $\sim$7.8~day oscillations occur below the first 
exponential decaycurve.

We also fitted the first 100 days of the {\it SAS-3} CSL-A outburst 
light curve with the model of Augusteijn, Kuulkers \&\ Shaham (1993), 
in order to determine when the tertiary and subsequent
maxima are expected in their model. The data is best fit
with the following parameters (see Augusteijn et al.\ [1993] for a description 
of these parameters): $\gamma^{-1} = 26$~days, $\beta^{-1} = 2.1$~days, 
$T = 49$~days and $\alpha\psi = 0.16$; the fit is shown in 
Fig.~\ref{a0620_allxray}. 
Note that $\gamma^{-1}$, $\beta^{-1}$ and $\alpha\psi$ are
comparable to those found for \astrobj{GS2000+25}{GS\,2000+25} 
(Augusteijn et al.\ 1993). 
Extrapolating the fit gives the following times of the expected tertiary,
quaternary and quinternary maxima: $\sim$JD\,2442739, $\sim$JD\,2442792 and 
$\sim$JD\,2442842, respectively.

Carpenter et al.\ (1976) noted flaring in the {\it Ariel~V} RMC light 
curve near JD\,2442643. This is consistent with the {\it Ariel~V} ASM 
measurements at JD\,2442644--645 which lie above that expected from a 
pure exponential decay of the data between JD\,2442658 and the secondary 
maximum (Kaluzienski et al.\ 1977; see also Fig.~\ref{a0620_detail}a). 
This may be due to a small bump in the 
light curve just after maximum, as sketched in the composite light 
curves by Carpenter et al.\ (1976) and White et al.\ (1984). 
Fig.~\ref{a0620_sas3}a reveals that the pure 
exponential decay started somewhat later at higher ($\gtrsim$6\,keV) energies,
which may indicate that the bump is more pronounced at these higher 
energies.

The {\it Ariel~V} RMC light curve shows a small glitch in its light curve
near JD\,2442654, as noted by Carpenter et al.\ (1976) and 
Pounds et al.\ (1977). This cannot be seen in the {\it SAS-3} CSL-A and 
CSL-B light curves. We note that the {\it SAS-3} CSL-A data as presented by
Matilsky et al.\ (1976) also showed an erratic decay up to the secondary
maximum, but this has been calibrated out in the improved analysis by Plaks
(1991), which is shown here. So, whether the small glitch in the {\it Ariel~V} 
RMC light curve is real or due to systematics in the data analysis is not 
clear.

\begin{figure}
\centerline{
\includegraphics*[bb = 66 345 477 701,width=10cm]{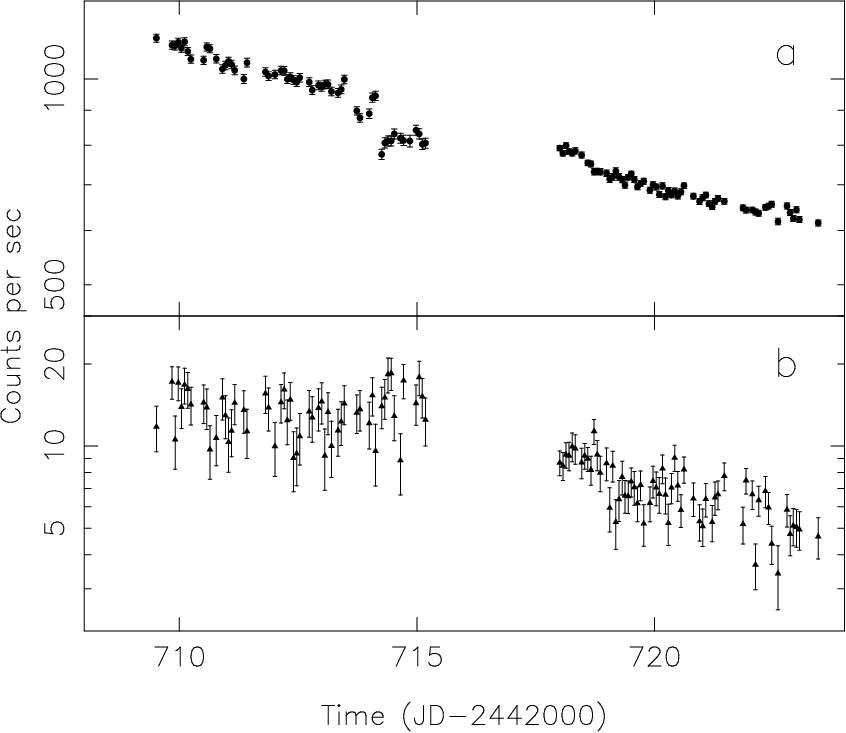}}
\caption{\small Blow-up of the {\it SAS-3} CSL-A (\textbf{a.}) and
CSL-B (\textbf{b.}) light curves around the time of the drop in intensity 
($\sim$JD\,2446715) in the {\it SAS-3} CSL-A light curve.}
\label{a0620_sas3_dip}
\end{figure}

A small sharp dip in the {\it Ariel~V} SSE light curve was noted by 
Pounds et al.\ (1977) with a minimum near JD\,24462730.8 (see also 
Fig.~\ref{a0620_detail}b) and lasted for about 2 days. 
Another small sharp dip, which 
has not been noted before, can be seen in the {\it SAS-3} CSL-A light 
curve which is in progress near JD\,2446714.5 and lasted for at least a day
(see Fig.~\ref{a0620_sas3_dip} for a blow-up of the {\it SAS-3} CSL-A and
{\it SAS-3} CSL-B lightcurves). 
This dip occurs within the uncertainties of the {\it Ariel~V} ASM light curve.
At higher energies ($\gtrsim$6\,keV), however, the 
{\it SAS-3} CSL-B (Fig.~\ref{a0620_sas3_dip}) and CSL-C light curves 
(Fig.~\ref{a0620_sas3}a) are more or less flat 
just before the dip and show a small increase, instead of a dip, which
causes a small increase of the hardness during that time
(Fig.~\ref{a0620_sas3}b). This means that during that time
the X-ray spectrum pivoted near 6\,keV with the low energy count rate
decreasing, while the high energy count rate was increasing.

Whereas the {\it SAS-3} CSL-A count rate gradually decreased from the 
secondary maximum until $\sim$2442780, the {\it SAS-3} CSL-B clearly shows 
a higher count rate than that extrapolated from the exponential decay after 
the secondary maximum. The {\it SAS-3} CSL-B intensity is consistent 
with being constant between JD\,24462747 and JD\,24462797. 
This results in a clear hardening of the source (Fig.~\ref{a0620_sas3}b; see 
also Plaks 1991) which has not been noted before. The hardening started 
somewhere between JD\,24462732 and JD\,24462747.
Note that the expected time of the tertiary maximum in the 
model of Augusteijn et al.\ (1993), i.e.\ JD\,24462739, lies within
this range. 

As noted by Kaluzienski et al.\ (1977) and Pounds et al.\ (1977) the 
minima in the {\it Ariel~V} SSE and ASM light curves between JD\,2442772 
and JD\,2442807 correspond to the minima which occur in the {\it SAS-3} 
CSL-A light curve. Note that the minima in the {\it Ariel~V} SSE light 
curve appear to be broader than those in the {\it SAS-3} CSL-A light 
curve. This is solely due to the sampling (see Watson 1982) of the 
{\it Ariel~V} SSE data at intervals of a few days during the trough. 
We note that short 
sharp decreases in the {\it Ariel~V} ASM light curves also appear near 
JD\,2442758, JD\,2442814 and JD\,2442854 (see 
also Fig.~\ref{a0620_detail}d). 

The broad tertiary maximum peaked around JD\,2442836 
(see also Fig.~\ref{a0620_detail}d). 
After JD\,2442859 \astrobj{A0620-00}{A\,0620$-$00} disappeared below 
the {\it Ariel~V} ASM instrument threshold (Kaluzienski et al.\ 1977).

\subsubsection{X-ray versus other wavelengths}\label{comparison}

\begin{figure}
\centerline{
\includegraphics*[bb = 66 109 572 605,width=12cm]{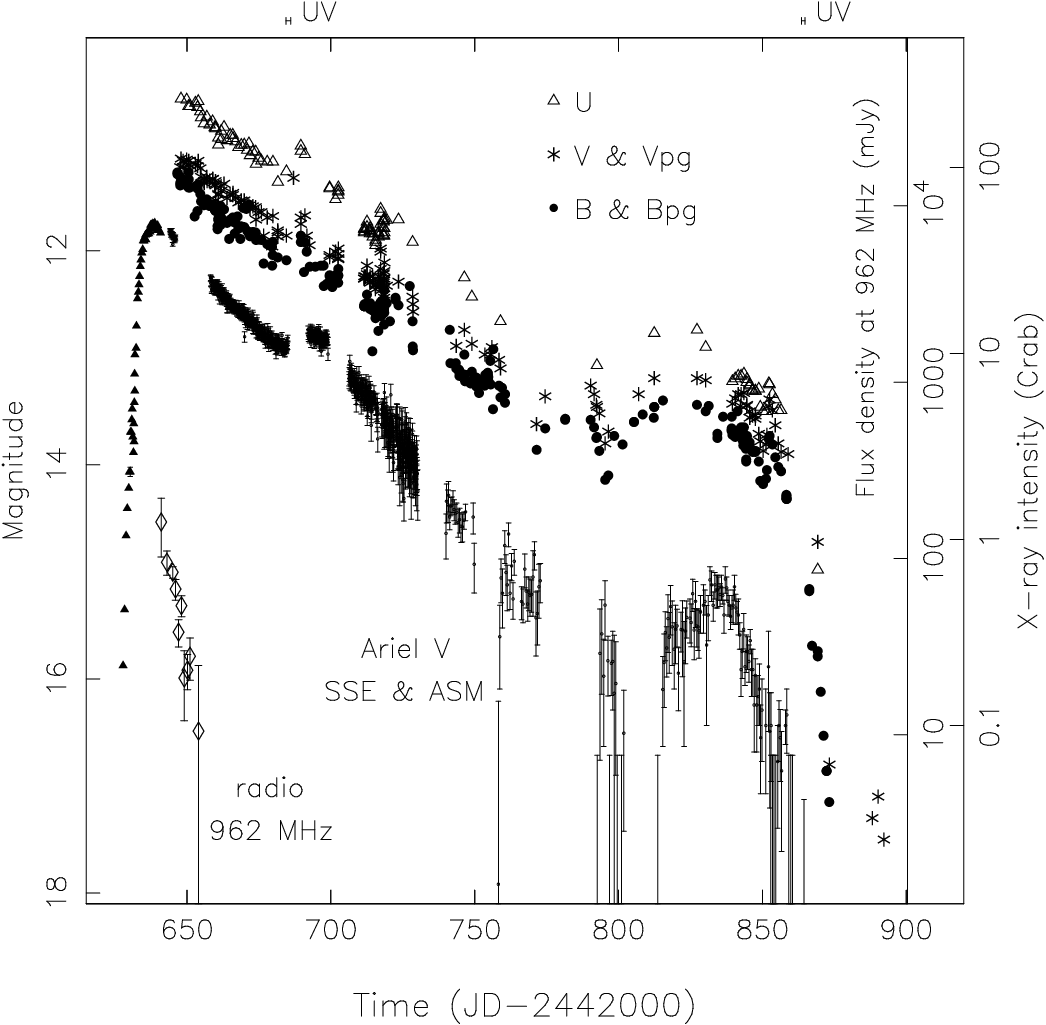}}
\caption{\small Overall outburst light curve in the optical (photographic and
photoelectric UBV data) and X-ray ({\it Ariel V} ASM and part of the SSE 
data) of \astrobj{A0620-00}{A\,0620$-$00}. The SSE light curve has 
been scaled (by a factor of 2.1) to match the exponential decay of the 
first $\sim$50~days of the ASM light curve. 
For reference we also included the radio light curve; the times of the 
{\it ANS} ultra-violet observations are indicated at the top.}
\label{a0620_ubv_X_radio_uv}
\end{figure}

\begin{figure}
\centerline{
\includegraphics*[bb = 28 43 587 747,angle=-90,width=14cm]{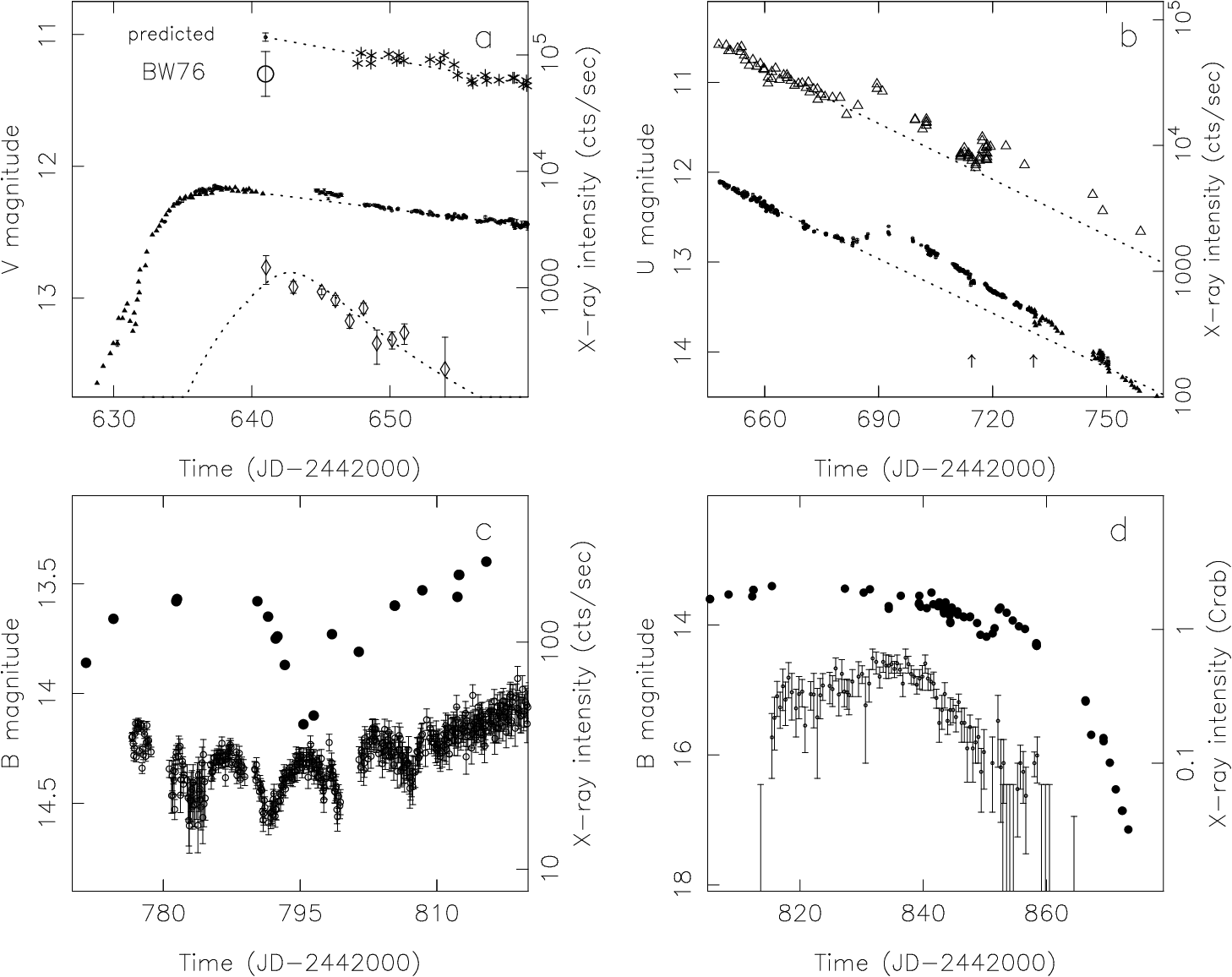}}
\caption*{\footnotesize \textbf{a.} The first $\sim$50~days of the outburst of 
\astrobj{A0620-00}{A\,0620$-$00}. 
Shown are the V band, {\it Ariel V} SSE and ASM, {\it SAS-3} CSL-A data, and 
the radio measurements. The {\it Ariel V} SSE and ASM light curves have been 
scaled to match the {\it SAS-3} CSL-A light curve. Plotted is also the 
optical estimate of \astrobj{A0620-00}{A\,0620$-$00} during
the optical discovery observation ($\circ$), and the expected value as 
extrapolated from the fit to the V band data ({\small $\bullet$}).
For the units of the radio data we refer to Fig.~\ref{a0620_ubv_X_radio_uv}.
Shown are the linear fit to the V band data between $\sim$JD\,2442648 and 
$\sim$JD\,2442682, the exponential fit to part of the 
first exponential decline (JD\,2442658--JD\,2442682) of {\it Ariel~V} ASM 
data and its extrapolation to the beginning and the end of the outburst,
and the synchroton radio source model fit to the radio data by 
Hjellming et al.\ (1988).
\textbf{b.} The exponential decay phase of the outburst of 
\astrobj{A0620-00}{A\,0620$-$00}. Shown 
are the U band measurements, and the X-ray ({\it Ariel V} SSE and {\it SAS-3}
CSL-A) data. The {\it Ariel V} SSE has been shifted to match the {\it SAS-3}
CSL-A light curve. The arrows indicate the times of the dips in the 
X-ray light curve. Fits to the first exponential decay in the U band and 
{\it SAS-3} CSL-A X-ray light curve and their extrapolation are plotted as
dotted lines.
\textbf{c.} The light curve during the trough of the outburst of 
\astrobj{A0620-00}{A\,0620$-$00}.
Shown are the B band measurements and the {\it SAS-3} CSL-A data.
\textbf{d.} The light curve at the end of the outburst of 
\astrobj{A0620-00}{A\,0620$-$00}.
Shown are the B band measurements and the {\it Ariel V} ASM data.
Symbols in the four figures are similar to those used in 
Figs.~\ref{a0620_allxray} and \ref{a0620_ubv_X_radio_uv}.} 
\label{a0620_detail}
\end{figure}

In Fig.~\ref{a0620_ubv_X_radio_uv} 
we present the outburst of \astrobj{A0620-00}{A\,0620$-$00} viewed over all 
wavelengths. This figure shows the optical (UBV) light curves, together with 
the X-ray outburst light curve from {\it Ariel V} ASM data and part 
(up to JD\,24462641) of the {\it Ariel V} SSE data.
For reference we also give the radio 
light curve at 962\,MHz as presented by Davis et al.\ (1975), and
at the top of the figure we have indicated the observation times of the 
{\it ANS} ultra-violet experiment (see Wu et al.\ 1976; 1983).
The radio outburst is clearly very short, i.e.\ a couple of 
weeks (see also Fig.~\ref{a0620_detail}a). 
The ultra-violet observations by Wu et al.\ (1976, near 
JD\,2442685) were obtained just at the beginning of the secondary 
maximum, whereas those reported by Wu et al.\ (1983, near 
JD\,2442864) were obtained during the first part of the final decay to 
quiescence (about a week after the source dropped below the detection 
limits of the {\it Ariel V} ASM\footnote{Kaluzienski et al.\ (1976) reported
an upper limit of $\sim$0.04~Crab near JD\,2442864.5 over a 0.5~day 
integration period.}). 
We note that Tsunemi et al.\ (1977) found that the very first decay seemed 
to be more rapid ($\sim$0.018~mag per day) than the rest of the main decay. 
Our data cover the same time base as theirs; we found, however, no clear 
evidence for different decay rates.

In Fig.~\ref{a0620_detail}a 
we zoom in on the first $\sim$50~days of the outburst.
As noted in the previous section, the X-ray maximum of the outburst was 
reached near JD\,2442638. From 
synchrotron radio source model fits (Fig.~\ref{a0620_detail}a; see Hjellming et 
al.\ 1988; Hjellming \&\ Han 1995) a maximum near 
JD\,2442642 is derived for the 962\,MHz radio data, while it is estimated to 
be $\sim$JD2442641 and $\sim$JD2442639.5 for data at 1.4\,GHz and 2.7\,GHz,
respectively. Although this suggests that the radio outburst was delayed with 
respect to the X-ray outburst, it is clear that the radio data are consistent 
with only a decay and that the maximum could have occurred earlier.
It may be worthwhile to note that the expected start of the
radio outburst in the synchrotron radio source 
model fit is $\sim$2442632, i.e.\ very close to the time of
the precursor in the {\it Ariel V} SSE data.

The optical counterpart of \astrobj{A0620-00}{A\,0620$-$00} was discovered by 
Boley et al.\ (1976) in the early morning of August 16 
(JD\,2442641).
Their measurements on this day and the following two days revealed 
the object to be of magnitude 12 or brighter. Since these measurements 
may give us an estimate of when the optical maximum was reached, we 
decided to study these observations more carefully.
We note that Ciatti et al.\ (1977) estimated the 
optical maximum to be near JD\,2442650, but they used a
$\sim$12th magnitude estimate by Boley et al.\ (1976). 

The observing conditions were very bad (it was $\sim$20~min before 
sunrise) and the telescope drive was not functioning in the contorted 
position Boley and Wolfson had to get it into
(Wolfson, 1997, private communication; see also 
Boley et al.\ 1976). But, examining their discovery image, an image tube 
photograph (Boley et al.\ 1976), one can still get a rough 
estimate of the actual magnitude during the discovery.
Comparing the magnitudes of the comparison stars in the frame with those
reported by Webbink (1978) one deduces that the image tube must correspond
more closely to V band than the B band.
By comparing the brightness of \astrobj{A0620-00}{A\,0620$-$00} with that of 
four comparison 
stars, one can derive ${\rm V} = 11.30 \pm 0.17$~mag, where the error is 
a standard deviation from estimates with respect to four 
comparison stars (Webbink 1997, private communication).

The estimate of the Boley et al.\ (1976) observation is also plotted in
Fig.~\ref{a0620_detail}a, 
together with the expected value (11.02$\pm$0.03) as extrapolated 
from a linear fit to the first decay phase in the V band. Within the 
uncertainties, the observed and expected magnitudes do not match. 
We note, however, that the bright sky and the high airmass also affects
our estimate from the Boley et al.\ (1976) observation 
and therefore its uncertainty may be even larger.
We cannot therefore say with certainty if the 
optical maximum was reached between Boley et al.'s observation
($\sim$JD\,2442641) and the start of the V band measurements 
($\sim$JD\,2442648), or e.g.\ that near maximum the optical shows a plateau.
If the former is true, then optical maximum was reached after the 
X-rays reached its maximum. In the latter case, the maximum would have been 
reached before $\sim$JD\,2442641. The B measurements start about one day 
earlier than the V band measurements and do not show evidence for flattening.
Also, a visual estimate at JD\,2442645.6 by Locher (1975)
of V$\sim$10.4 is available (see also Fig.~\ref{a0620_quiescence_b_v}); 
this estimate is, however, far from both the 
estimate of Boley et al.'s observation and that expected from 
extrapolation from the first exponential decay in V, and we suspect 
therefore that Locher's measurement has been affected by systematic 
errors in the magnitudes of comparison stars 
(Section~\ref{optical_observations}).

Near the time of the X-ray secondary maximum, there is also a secondary 
maximum in the UBV outburst light curve (Figs.~\ref{a0620_ubv_X_radio_uv}
and \ref{a0620_detail}b). 
The rise to secondary maximum starts more or less simultaneously
in the optical and X-ray (see also Section~\ref{ir}). 
However, the optical peak was reached near JD\,2442687, whereas 
in X-rays it was reached near JD\,2442692 (Section~\ref{x-ray}).
This suggests a delay in X-rays with respect to the optical of $\sim$5~days.

The second exponential decay phase in X-rays is 
faster than the first exponential decay phase (Section~\ref{x-ray}) 
and seems to be rather smooth
before and after the secondary maximum. The second exponential decay in 
the U band is even faster (see Fig.~\ref{a0620_detail}b), and therefore 
the optical catches up with that expected
from the first exponential decay earlier with respect to that in X-rays
(see also Section~\ref{ir}).

About 30~days after the secondary maximum the optical UBV shows another
local maximum (possibly preceded by a depression), 
which is {\it not} seen in the 
X-ray light curve. It is clearly visible in the 
U band data (Fig.~\ref{a0620_detail}b). 
This feature in the light curve is the modulation 
reported by Chevalier et al.\ (1980); we will denote this as an
`intermediate maximum'. The depression just before the 
peak of the `intermediate maximum' occurs exactly at the same time as 
the small X-ray dip in the {\it SAS-3} CSL-A light curve. 
Unfortunately, no simultaneous optical data are available
during the other small X-ray dip, near JD\,24462731 as observed with the 
{\it Ariel V} SSE (Section~\ref{x-ray}).

Two minima during the trough in the B and V light curve 
(Fig.~\ref{a0620_ubv_X_radio_uv}) are seen near 
JD\,2442771 and JD\,2442796 (see also Robertson et al.\ 1976). 
In Fig.~\ref{a0620_detail}c 
we zoom in on the trough of the outburst, just before the 
large bump at the end of the outburst light curve. The $\sim$7.8~day X-ray
oscillations are clearly visible. The optical also shows evidence for 
minima and maxima during the same part of the outburst and are most 
probably related to the X-ray oscillations (see already 
Section~\ref{a0620_intro}). 
Fig.~\ref{a0620_detail}c shows clearly that the optical and X-ray are 
almost anti-correlated, as can especially
be seen between $\sim$JD\,2442790 and $\sim$JD\,2442797. If the optical 
tracks the X-rays, this means that the X-rays are delayed by 
$\sim$4.5~days with respect to the optical or that the optical is 
delayed by $\sim$4~days with respect to the X-rays.

The large bump at the end of the light curve is broader in the optical
compared to the X-rays. As can be seen in Fig.~\ref{a0620_detail}d 
and shown by Chen et al.\ (1993), the broad tertiary maximum started 
probably earlier by about $\sim$15~days in the optical than in X-rays. 
Also, the peak of the optical tertiary maximum was reached somewhere
between $\sim$JD\,2442816 and $\sim$JD\,2442827. In X-rays this peak was 
reached around JD\,2442836 (Section~\ref{x-ray}), suggesting a delay 
between the optical and X-ray peak of $\sim$10--20~days.
At the end of the outburst, just before the final decay, there is a dip 
followed by an increase in brightness of $\sim$0.5 magnitudes, resulting in 
another intermediate maximum (at $\sim$JD\,2442853), in the optical light 
curve. This is the `eclipse-like feature' of 
Chevalier et al.\ (1980), see Section~\ref{a0620_intro}. The {\it Ariel V} 
ASM light curve shows evidence for a similar kind of dip: at the end of 
the X-ray decay \astrobj{A0620-00}{A\,0620$-$00} rises above the detection 
limits near JD\,2442855 and drops below the detection limits again 
$\sim$4~days later. If this so-called `hiccup' in X-rays is related to that 
in the optical, it also suggests a delay between the optical and X-ray of 
$\sim$5~days.

\subsubsection{Infrared}\label{ir}

\begin{figure}
\centerline{
\includegraphics*[bb = 78 80 535 513,angle=-90,width=10cm]{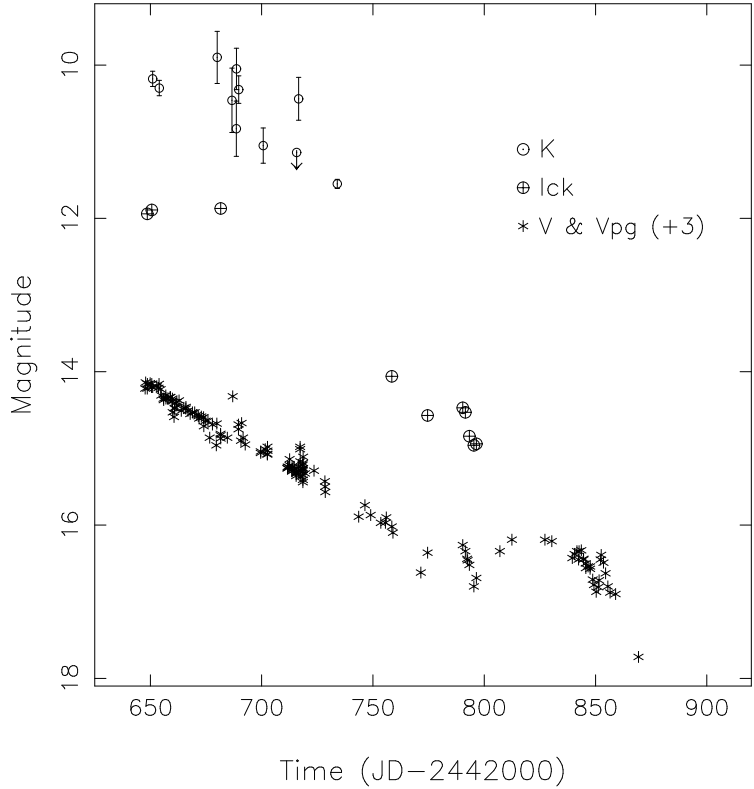}}
\caption{\small Infrared measurements (I$_{\rm ck}$ and K, see text) during the 
\astrobj{A0620-00}{A\,0620$-$00} outburst. For comparision we also show the V 
band data shifted 
by 3 magnitudes. Errors on the I$_{\rm ck}$ have not been given in the 
literature.}
\label{a0620_v_ir}
\end{figure}

In Fig.~\ref{a0620_v_ir} we show the Cape-Kron I (I$_{\rm ck}$, 
Robertson et al.\ 1976) 
and K band (Jameson 1975; Kleinmann et al.\ 1976; 
Citterio et al.\ 1976; Szkody 1977) measurements 
during the outburst of \astrobj{A0620-00}{A\,0620$-$00}, together 
with the V band light curve (shifted by 3 magnitudes). Data at other infrared 
wavelengths are available (e.g.\ R$_{\rm ck}$, J, and H), but we did not 
plot them, since their 
number of measurements is less than that of I$_{\rm ck}$ and K band data, and 
do not reveal additional information.

The K band measurements between JD\,2442680 and JD\,2442690 show some variations
in excess of that expected from a smooth decay, which is near the optical
(and X-ray) secondary maximum. We note that the K band measurement
near JD\,2442680 lies above that expected, assuming a decay rate similar to the
V band measurements from the beginning of the outburst. Similarly, the 
I$_{\rm ck}$ measurement near JD\,2442682 lies higher than expected from 
a smooth decay with a rate similar to the V band. From the 
optical data we infer that the rise to secondary maximum starts between 
JD\,2442685 and JD\,2442687; whereas this occurs near JD\,2442684 for the
X-rays. This may indicate that in the infrared activity already started 
before that in the optical and X-rays. 

As already noted by (Citterio et al.\ 1976) the two K band measurements near
JD\,2442716 show evidence for a sudden increase in the K band magnitude within
a day. This occurs exactly near the time of the `intermediate maximum'
(see Section~\ref{comparison}). We note, however, that their
simultaneous H band measurements did not show significant changes.

During the trough of the outburst the I$_{\rm ck}$ measurements follow 
the behaviour seen in the V band.

From Fig.~\ref{a0620_v_ir} it also becomes clear that in the V band 
the secondary maximum near JD\,2442687 and the intermediate maximum near
JD\,2442716 are much shorter lived than the secondary maximum in 
X-rays (Section~\ref{x-ray}), i.e.\ 
within $\sim$5~days the V band follows again 
the decay as extrapolated from the first exponential decay.

\subsubsection{Optical colour behaviour}\label{colour}

\begin{figure}
\centerline{
\includegraphics*[bb = 66 109 572 701,width=12cm]{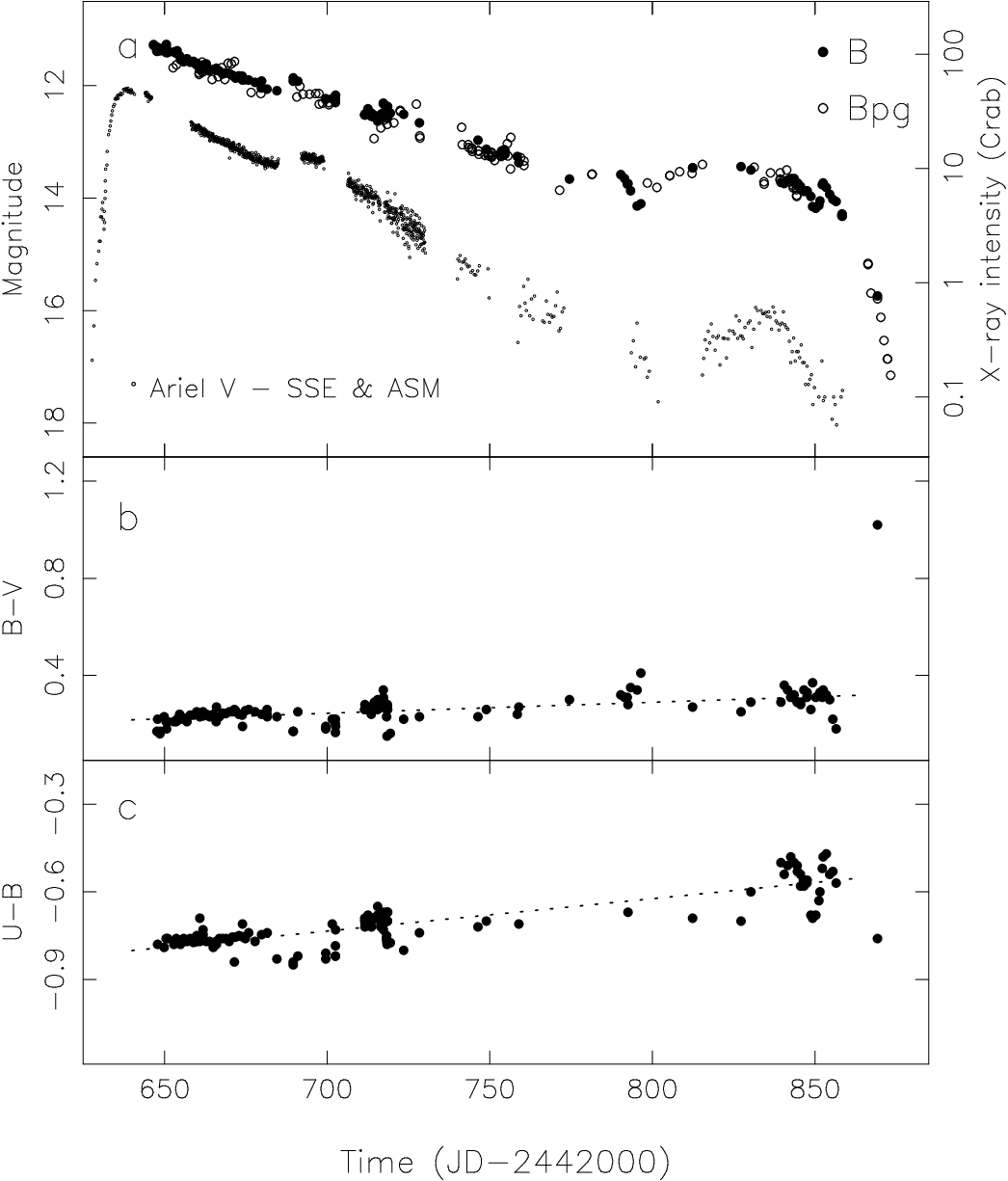}}
\caption{\small \textbf{a.} Photoelectric ($\bullet$) and photographic ($\circ$) B
magnitude measurements during the outburst of 
\astrobj{A0620-00}{A\,0620$-$00}. For reference we 
give the {\it Ariel V} SSE and ASM outburst light curve
(see also Fig.~4). Photoelectric
B$-$V (\textbf{b.}) and U$-$B (\textbf{c.}) measurements of 
\astrobj{A0620-00}{A\,0620$-$00}.}
\label{a0620_b_X_bmv_umb}
\end{figure}

Following Webbink (1978), but now with additional data, we plot the B 
magnitude during the outburst, together with the photoelectric 
B$-$V and U$-$B measurements. We have also plotted the linear fits to the 
B$-$V and U$-$B data in order to clarify changes in these measurements.
Lloyd et al.\ (1977, nearly simultaneously photographic BV data) and 
Lyutyi (1976, simultaneous photoelectric BV) concluded from their data 
sets that B$-$V stayed roughly constant during the whole outburst. The same 
was concluded for the photoelectric U$-$B measurements by Lyutyi (1976).
Duerbeck \&\ Walter (1976), however, reported a slight enhancement in the 
photoelectric B$-$V and U$-$B measurements during the first exponential decay.
The fit shows that, apart from small deviations, B$-$V slowly 
increased from $\sim$0.22 to $\sim$0.31 from the beginning of the outburst to 
just before the final decline. The change in U$-$B is larger; it increased 
from about $-$0.79 in the beginning of the outburst to about $-$0.56 just 
before the final decline. Deviations in B$-$V and U$-$B occur between 
$\sim$JD\,2442684 and JD\,2442700 (i.e.\ during the secondary maximum), 
after $\sim$JD\,2442718 (just after the `intermediate maximum'),
during the trough near $\sim$JD\,2442795 (see also Robertson et al.\ 1976), and
at the end of the tertiary maximum, i.e.\ around $\sim$JD\,2442850
(near the `eclipse-like feature').

At the final decline to quiescence photoelectric UBV measurements 
revealed a different behaviour as compared to the rest of the outburst
(see Lyutyi 1976; Lyutyi \&\ Shugarov 1979): between 
JD\,2442856.5 and JD\,2442869.3 B$-$V was already close to its quiescent value 
($\sim$1.3, e.g.\ Murdin et al.\ 1980), while U$-$B was still at an outburst 
value of $\sim$$-$0.76
(quiescence U$-$B is $\sim$0.5, e.g.\ Lyutyi \&\ Shugarov 1979). 
This indicates that in U and B the flux
dropped faster to quiescence than in V, as can be seen in 
Fig.~\ref{a0620_ubv_X_radio_uv}.

\subsection{Quiescence}\label{quiescence}

\begin{figure}
\centerline{
\includegraphics*[bb = 46 208 562 739,width=12cm]{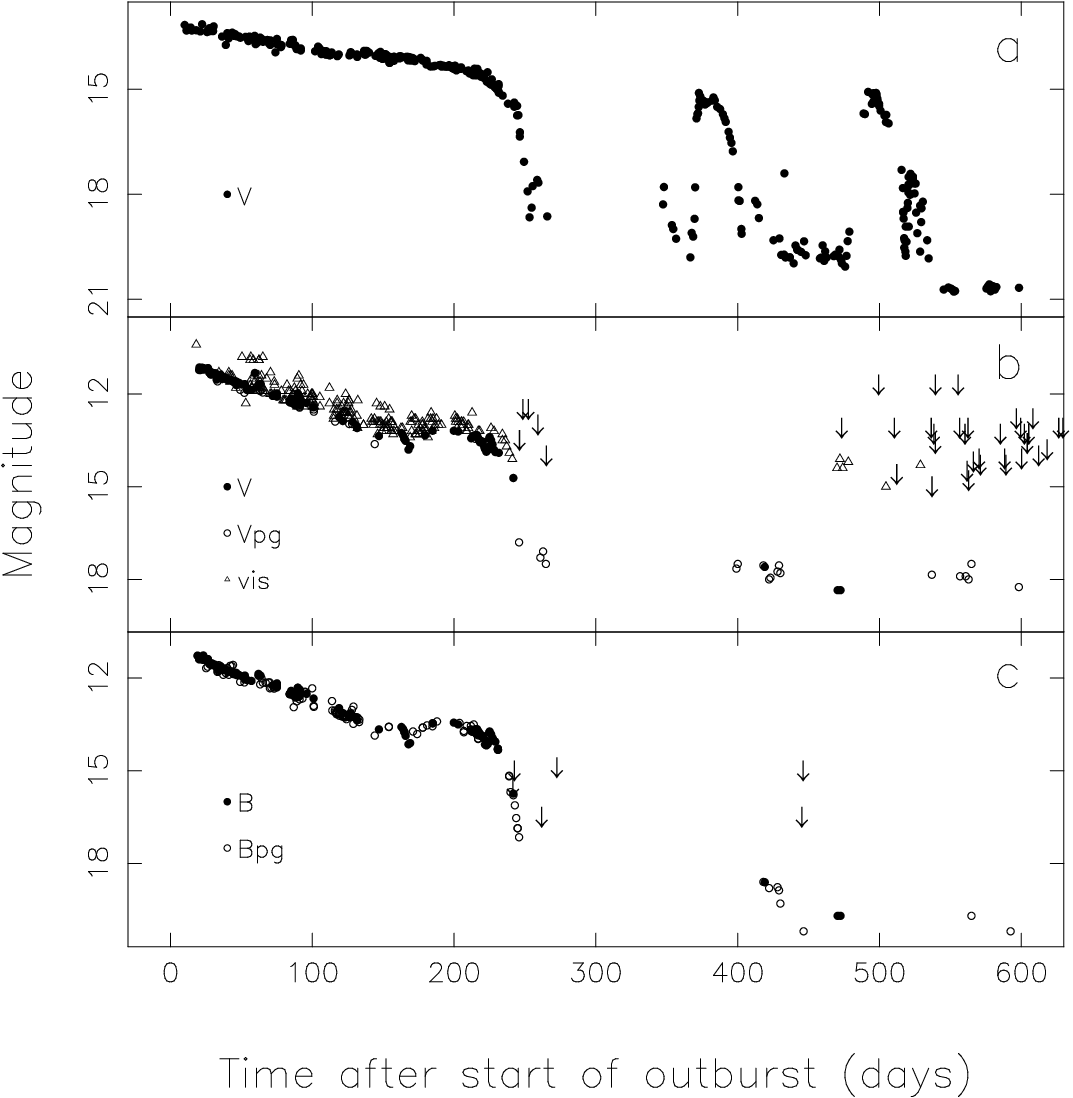}}
\caption*{\small \textbf{a.} V band light curve of \astrobj{GROJ0422+32}{GRO\,J0422+32}. 
Data are from 
Castro-Tirado et al.\ 1993, Mineshige 1994, 
Bartolini et al.\ 1994, Callanan et al.\ 1995, 
Chevalier \&\ Ilovaisky 1995, Kato, Mineshige \&\ Hirata 1995,
King, Harrison \&\ McNamara 1996, Goranskii et al.\ 1996, and
measurements reported in IAU Circulars, see Chevalier \&\ Ilovaisky 1995.
Visual and V band (\textbf{b.}), and B band (\textbf{c.}) light curves 
of \astrobj{A0620-00}{A\,0620$-$00}. Arrows in \textbf{b.} and \textbf{c.} 
denote lower limits on the magnitude from visual and photographic B 
measurements, respectively.}
\label{a0620_quiescence_b_v}
\end{figure}

Extensive optical monitoring of the SXT \astrobj{GROJ0422+32}{GRO\,J0422+32} 
(X-ray Nova Per 1992, \astrobj{V518Per}{V518\,Per}) after the main outburst 
has led to the detection of `mini-outbursts' or rebrightenings
(Callanan et al.\ 1995; Chevalier \&\ Ilovaisky 1995; see also
Shrader et al.\ 1997; Castro-Tirado, Ortiz \&\ Callego 1997). 
The only other SXT seen to exhibit
`mini-outbursts' is \astrobj{GRS1009-45}{GRS\,1009$-$45} (X-ray Nova Vel 1993, 
\astrobj{MMVel}{MM\,Vel}; Bailyn \&\
Orosz 1995; see Fig.~9).
In Fig.~\ref{a0620_quiescence_b_v} 
we show the V band optical light curve of 
\astrobj{GROJ0422+32}{GRO\,J0422+32} and 
the V and B band light curves of \astrobj{A0620-00}{A\,0620$-$00}
on the same time scale. The main outbursts of 
\astrobj{A0620-00}{A\,0620$-$00} and \astrobj{GROJ0422+32}{GRO\,J0422+32}
have a very similar duration. It is therefore interesting to investigate any
possible `mini-outbursts' in \astrobj{A0620-00}{A\,0620$-$00}, which 
might have been unnoticed.

There are several visual reports 
by Duruy (1976) around 470 and 530 days after the start of the outburst 
and an estimate by Cragg $\sim$505~days after the start of 
the outburst (see Bateson 1976), which indicate that 
\astrobj{A0620-00}{A\,0620$-$00} was seen $\sim$4--5 magnitudes above its 
quiescent level of
$\sim$18~mag. However, measurements done by Oke (1977) 472 and 474
days after the start of the outburst showed the system to be in quiescence, 
almost simultaneous with some of Duruy's estimates. This, therefore, casts some 
doubt on the visual estimates, and the existence of any possible 
rebrightening\footnote{Duruy used 
an AAVSO preliminary chart during outburst, whereas after the final decline 
of the system, he used a personal sequence. Like all the visual observers, 
even the most experienced, Duruy may well have made a mistake, but he never 
corrected these observations after their publication in the AFOEV bulletin 
(Schweitzer 1997, private communication). We note that Duruy later also 
reported several upper limits, which means he at least observed a 
variable star.}.

\subsection{A comparison with other short period SXTs}\label{sxts}

In Fig.~9 
we show the compilation of optical and X-ray observations
of the short period SXTs \astrobj{GROJ0422+32}{GRO\,J0422+32}, 
\astrobj{GRS1009-45}{GRS\,1009$-$45}, \astrobj{A0620-00}{A\,0620$-$00}, 
\astrobj{GS2000+25}{GS\,2000+25} (X-ray Nova Vul 1988, 
\astrobj{QZVul}{QZ\,Vul}) and \astrobj{GS1124-68}{GS\,1124$-$68} 
(X-ray Nova Mus 1991, \astrobj{GUMus}{GU\,Mus})\footnote{We note that 
West (1991) erroneously reports on an 
observation obtained on Jan.\ 13.25~UT 1988 (JD\,2447269.75) by G.~Pizarro, 
where the counterpart was found at 17--18~mag; this magnitude estimate is in 
fact obtained from an ESO Schmidt plate on Jan.~29 1976 taken by G.~Pizarro
(see Della Valle, Jarvis \&\ West 1991a; 1991b).}, 
where the orbital period increases from \astrobj{GROJ0422+32}{GRO\,J0422+32} 
($\sim$5.1~hr) to \astrobj{GS1124-68}{GS\,1124$-$68} ($\sim$10.4~hr), 
see e.g.\ Shahbaz \&\ Kuulkers (1998).
This figure shows that apart from secondary and tertiary maxima, which 
have been previously reported in X-ray and optical (e.g.\ Chen et al.\ 1997,
and references therein), all the outbursts appear to last for $\sim$250 days.
Also, for \astrobj{A0620-00}{A\,0620$-$00}, \astrobj{GS2000+25}{GS\,2000+25} 
and \astrobj{GS1124-68}{GS\,1124$-$68} both the optical and 
X-ray show drops in the flux away from the expected exponential decay
around $\sim$150~days after the start of the X-ray outburst, followed
by the tertiary maximum. In the optical there is a drop in magnitude in 
\astrobj{GRS1009-45}{GRS\,1009$-$45} $\sim$150~days after the start of its 
outburst, which has previously been related to one of its mini-outbursts 
(Bailyn \&\ Orosz 1995). A sudden fading of the optical flux of 
\astrobj{GROJ0422+32}{GRO\,J0422+32} 
has been reported by Bartolini et al.\ (1994), which occurred 
$\sim$148 days after the outburst (not plotted in
Fig.~9). Optical data taken shortly thereafter 
(Kato et al.\ 1995) indicate that this drop lasted at least 
$\sim$0.42 days. This all suggests that the morphology of the outbursts 
in the optical and X-rays of these SXTs is very similar.
Similar behaviour may be seen in the 1971 outburst light curve of
\astrobj{4U1543-47}{4U\,1543$-$47} as recorded by {\it UHURU}, {\it OSO-7} and 
{\it Vela} (Li, Sprott \&\ Clark 1976; see also Chen et al.\ 1997):
$\sim$140~days after the maximum outburst (which showed an exponential 
decay together with a secondary maximum) the X-ray intensity starts to 
show various deep drops below the extrapolated exponential decay.

\begin{figure}
\parbox{7.2cm}{
\centerline{
\vspace{0.5cm}
\includegraphics*[bb = 66 35 403 769,width=7.2cm]{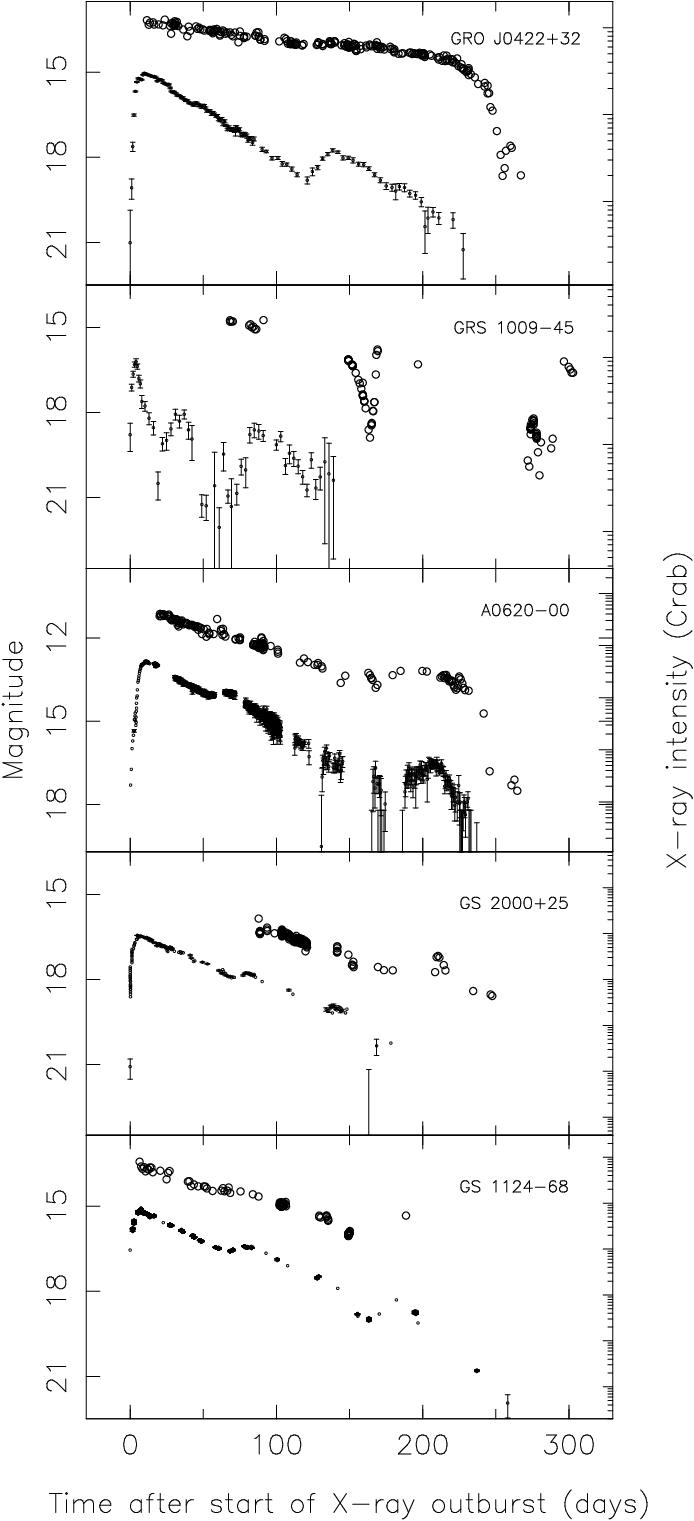}
} 
} 
\hspace{0.3cm}
\parbox{6.5cm}{
\caption*{\footnotesize Optical ($\circ$) outburst light curves of \astrobj{GROJ0422+32}{GRO\,J0422+32} 
(V), \astrobj{GRS1009-45}{GRS\,1009$-$45} (V), 
\astrobj{A0620-00}{A\,0620$-$00} (V), 
\astrobj{GS2000+25}{GS\,2000+25} (R), and 
\astrobj{GS1124-68}{GS\,1124$-$68} (V) and the corresponding X-ray light 
curves as a function of
the start of the X-ray outburst. For references to the optical data of 
\astrobj{GROJ0422+32}{GRO\,J0422+32}, see Fig.~\ref{a0620_quiescence_b_v}.
The optical V band data of \astrobj{GRS1009-45}{GRS\,1009$-$45} are from 
Bailyn \&\ Orosz (1995), Masetti, Bianchini \&\ Della Valle 
(1997) and Della Valle et al.\ (1997), while those of 
\astrobj{GS1124-68}{GS\,1124$-$68} are from Bailyn (1992), 
King, Harrison \&\ McNamara (1996), and 
Della Valle, Masetti \&\ Bianchini (1998). The R band data of 
\astrobj{GS2000+25}{GS\,2000+25} are from Borisov et al.\ (1989) 
[see also Karitskaya (1989)] and Charles et al.\ (1991).
We have added a constant of 0.8 to the data of Borisov et al.\ 
(1989) so that they agree with those of Charles et al.\ (1991).
We have also included the V band data (near 210 days after start of the 
outburst of Chevalier \&\ Ilovaisky (1990) and applied a colour 
correction of $\sim$1 (see Charles et al.\ 1991).
The X-ray data are taken from Chen et al.\ (1997). ${\rm T}=0$ corresponds to
2448838.5, 2449241.6, 2442627.5, 2447273.8 and 2448265.3 for 
\astrobj{GROJ0422+32}{GRO\,J0422+32}, \astrobj{GRS1009-45}{GRS\,1009$-$45}, 
\astrobj{A0620-00}{A\,0620$-$00}, \astrobj{GS2000+25}{GS\,2000+25}, and 
\astrobj{GS1124-68}{GS\,1124$-$68}, respectively.}
} 
\label{nper_nvel_nmon_rnvul_nmus}
\end{figure}

For \astrobj{GRS1009-45}{GRS\,1009$-$45} and 
\astrobj{GROJ0422+32}{GRO\,J0422+32} no soft
X-ray information (typically 2--20\,keV) is available, so it is difficult 
to compare them with the X-ray light curves of the other SXTs, although we 
note that the hard X-ray light curves of \astrobj{A0620-00}{A\,0620$-$00} 
(this paper) and \astrobj{GS1124-68}{GS\,1124$-$68} 
(Ebisawa et al.\ 1994) become more irregular and
less compatible with a so-called `fast-rise exponential decay' (FRED) light
curve, which seems to be similar to \astrobj{GRS1009-45}{GRS\,1009$-$45}. 
However, this does not apply to \astrobj{GROJ0422+32}{GRO\,J0422+32} 
(Fig.~9).

\subsection{The outbursts of A\,0620$-$00 and AL\,Com}\label{AL_Com}

It was quickly recognized that the optical amplitude of the outburst
of \astrobj{A0620-00}{A\,0620$-$00} was comparable to that of dwarf novae such 
as \astrobj{VYAqr}{VY\,Aqr}, \astrobj{UZBoo}{UZ\,Boo}, 
\astrobj{ALCom}{AL\,Com} and \astrobj{WZSge}{WZ\,Sge} 
(e.g.\ Eachus et al.\ 1976; Kholopov \&\ Efremov 1976; 
Richter 1986). Also the recurrence 
times of the outbursts of these systems were found to be similar to 
e.g.\ \astrobj{A0620-00}{A\,0620$-$00} and \astrobj{AqlX-1}{Aql\,X-1} 
(\astrobj{V1333Aql}{V1333\,Aql}), see e.g.\ Richter (1986).
In fact, all short period SXTs have large optical amplitude outbursts
(see e.g.\ Tanaka \&\ Lewin 1995; Shahbaz \&\ Kuulkers 1998; see also 
Fig.~9).
The dwarf novae mentioned above all belong to a subclass of the
SU\,UMa stars, called 
`Tremendous Outburst Amplitude Dwarf novae' 
(TOADs, Howell, Szkody \&\ Cannizzo 1995), also referred to
as \astrobj{WZSge}{WZ\,Sge} stars 
(e.g.\ Warner 1995; see a discussion in Patterson et al.\ 1996). 

\begin{figure}
\centerline{
\includegraphics*[bb = 28 71 579 672,width=11cm]{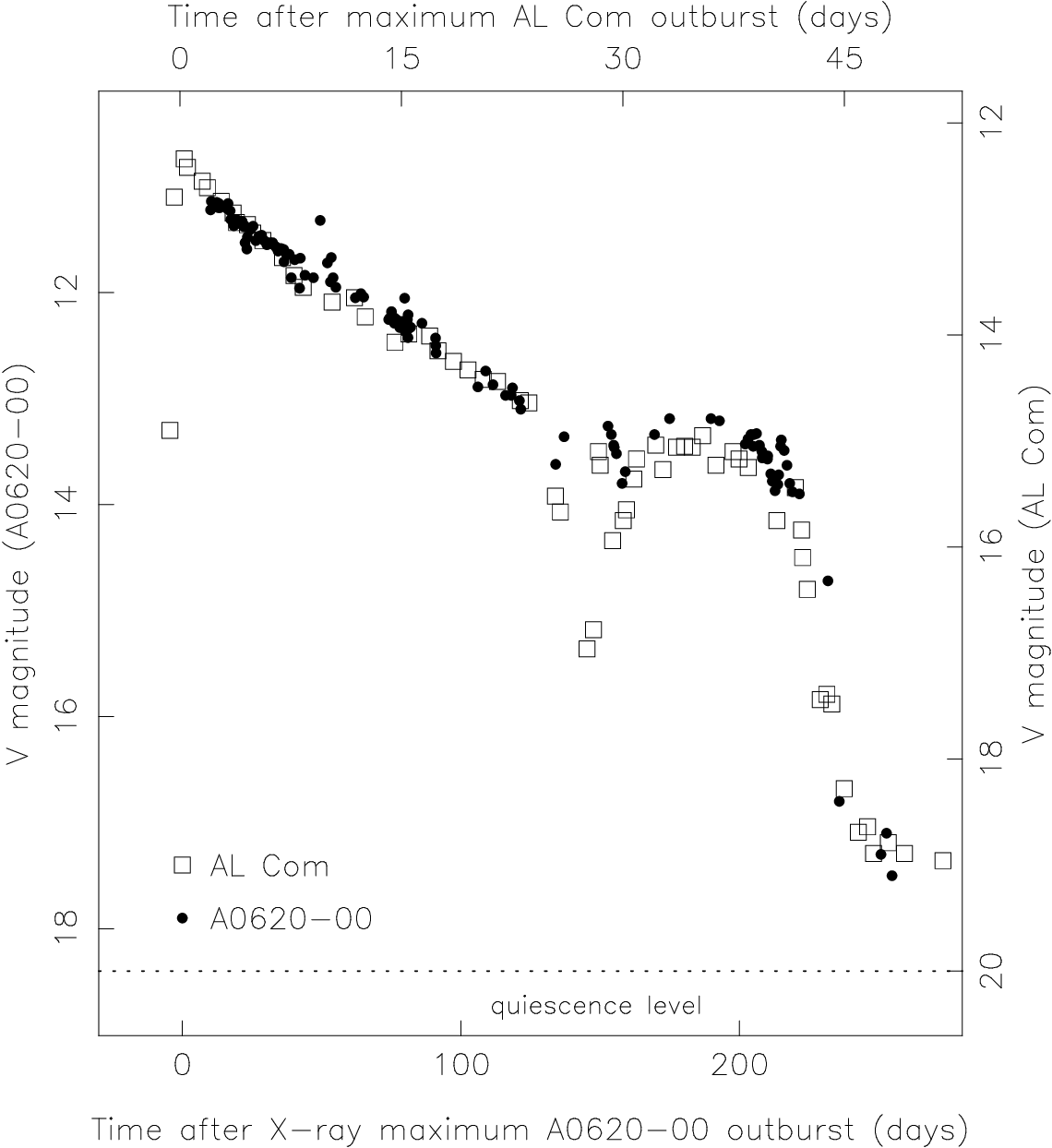}}
\caption{\small V band outburst light curve of \astrobj{A0620-00}{A\,0620$-$00} and 
\astrobj{ALCom}{AL\,Com} plotted on top of each other. The data of 
\astrobj{ALCom}{AL\,Com} are taken from
Patterson et al.\ (1996); most errors are smaller that the symbol size. 
The x-axes of \astrobj{ALCom}{AL\,Com} has been scaled 
(factor $\sim$5.3) to match the outburst length of 
\astrobj{A0620-00}{A\,0620$-$00}; 
the y-axes have only been shifted in magnitude to match the
quiescent level of both sources.}
\label{a0620_alcom}
\end{figure}

The last outburst of \astrobj{ALCom}{AL\,Com}, in 1995, has been fairly well 
covered by various observers (Howell et al.\ 1996; 
Patterson et al.\ 1996; Nogami et al.\ 1997).
In Fig.~\ref{a0620_alcom} we have superposed the outburst light curve
of \astrobj{ALCom}{AL\,Com} on that of \astrobj{A0620-00}{A\,0620$-$00}.
Note that the x-axes of \astrobj{ALCom}{AL\,Com} and 
\astrobj{A0620-00}{A\,0620$-$00} scale by a factor $\sim$5.3, while the 
y-axis scales are the same (but have been shifted in magnitude).
This shows that the outburst amplitudes are similar and that the shape of the 
outburst of \astrobj{ALCom}{AL\,Com} is almost a carbon copy of that of 
\astrobj{A0620-00}{A\,0620$-$00}, including the two 
enhancements during the exponential decline, the drops in intensity at the 
end of this decline, the final bump at the end of the outburst and the small 
local maximum before the final fast decay to $\sim$1 magnitude above their 
quiescence values. 

\section{Discussion}\label{discussion}

In this paper we have compared the outburst light curves at various 
wavelengths of \astrobj{A0620-00}{A\,0620$-$00}. Although we confirm several 
of the features reported earlier, we found various new features not seen before.

\subsection{X-ray delay}

During several stages of the outburst we found evidence for delays between
the optical and X-ray. The peak of the secondary maximum was reached
$\sim$5 days earlier in the optical. The infra-red activity associated 
with the secondary maximum started even a few days earlier with respect to the 
optical and X-ray. During the trough 
of the outburst, drops in intensity are consistent with a similar
delay of $\sim$5~days. Near the end of the outburst a `hiccup' occurred in 
the optical, which was followed $\sim$5 days later in the X-rays.
The broad tertiary maximum started about two weeks earlier in the optical
with respect to X-rays; also the peak of this broad maximum was consistent
with a similar delay. 

A delay between the optical and X-ray was also reported for 
\astrobj{GS1124-68}{GS\,1124$-$68} 
by Ebisawa et al.\ (1994). They found that the peak 
of the X-ray outburst was reached $\sim$4~days later than the 
peak of the optical outburst, i.e.\ a very similar timescale to
\astrobj{A0620-00}{A\,0620$-$00}. We note that we cannot say with certainty 
if a similar delay is present at the peak of the outburst of 
\astrobj{A0620-00}{A\,0620$-$00}.
More recently, a delay of $\sim$6~days between the optical and X-ray was found 
at the start of the outburst of the superluminal SXT 
\astrobj{GROJ1655-40}{GRO\,J1655$-$40} (Orosz et al.\ 1997). 

The delays seen between the optical and X-rays indicates that most of 
the processes associated with the outburst occur in the outer parts of the 
accretion disk. The outburst starts somewhere in the outer disk, 
because of the presence of an inner advection-dominated accretion flow (ADAF)
which prevents forming an inner disk region during quiescence
(see Narayan, McClintock \&\ Yi 1996; 
Lasota, Narayan \&\ Yi 1996; Narayan, Barret \&\ McClintock 1997; 
Hameury et al.\ 1997). Additional X-rays appear only when the 
disk is able to extend inwards and therefore causes a delay. 
Renewed activity in the outer parts of the disk 
(e.g.\ Ichikawa \&\ Osaki 1994; King \&\ Ritter 1998) 
may explain the delays seen between the optical (and infra-red) and X-rays
(not all material has been accreted and it may already have returned to its 
cool state, or fresh material has been provided by the companion star). 
Such renewed activity might be induced by the intense irradiation causing the 
outer parts of the accretion disk to return to a high state again
(e.g.\ King \&\ Ritter 1998).
We note that activity in the outer disk is also revealed by the onset of 
superhumps in the SXTs, which always occur just after the secondary maximum 
(see O'Donoghue \&\ Charles 1996). The fact that we see optical 
superhumps despite the luminous accretion disk has also been attributed to 
irradiation (O'Donoghue \&\ Charles 1996, see also Billington et al.\ 
1996).

\subsection{Outburst decay and local X-ray maxima}

As already noted, the first exponential decay phase of the outburst
of \astrobj{A0620-00}{A\,0620$-$00} is slower than its second exponential 
decay phase. This means that in the case of \astrobj{A0620-00}{A\,0620$-$00} 
the secondary maximum is
a temporary enhancement on the first exponential X-ray decay 
light curve. Also in the optical the secondary (and `intermediate')
maxima decay more rapidly in order to resume the same exponential
decay as seen before the secondary maximum. In the optical
this occurs much earlier than in X-rays ($\sim$1 week in the optical 
versus $\sim$2 months in X-rays).
For the other (short-period) SXTs with comparable outburst light 
curves this seems not to be the case, i.e.\
the decay timescales before and after the secondary maxima are
similar (see e.g.\ Augusteijn et al.\ 1993; Chen et al.\ 1993; 1997;
see also Fig.~9).

It is also evident from the overall outburst light curve of
\astrobj{A0620-00}{A\,0620$-$00} that the 
time difference between the main maximum and the secondary maximum and
that of the secondary maximum and tertiary maximum are not of the same 
order, as seems to be the case for \astrobj{GS2000+25}{GS\,2000+25} 
(Augusteijn et al.\ 1993, see also Chen et al.\ 1993).
However, the inferred third maximum by Augusteijn et al.\ (1993) is not the 
real tertiary maximum; the tertiary maximum is probably reached
$\sim$200~days after the peak of the outburst 
(see Kitamoto et al.\ 1992; Terada et al.\ 1998; see also
Tanaka \&\ Shibazaki 1996).
Also for \astrobj{GS1124-68}{GS\,1124$-$68} the time differences between the 
different maxima are not similar (see Ebisawa et al.\ 1994; see also
Fig.~9), and its tertiary maximum is $\sim$200~days after peak of the outburst
(see Ebisawa et al.\ 1994; see also Tanaka \&\ Shibazaki 1996), i.e.\ 
similar to A\,0620$-$00 and GS\,2000+25. 
It has recently been shown that the time of the secondary maximum 
in SXTs is related to the viscous time scale of an irradiated disk
(Shahbaz et al.\ 1998b; see also King \&\ Ritter 1998).
The time of the tertiary maximum seems to be unrelated to this.

It is interesting to note, however, that the times of the expected tertiary 
maxima of the outbursts of \astrobj{A0620-00}{A\,0620$-$00}, 
\astrobj{GS2000+25}{GS\,2000+25} and \astrobj{GS1124-68}{GS\,1124$-$68} in 
the model of Augusteijn et al.\ (1993) are consistent with the times of start of
the spectral hardening (see also next Section; 
\astrobj{A0620-00}{A\,0620$-$00}: this paper;
\astrobj{GS2000+25}{GS\,2000+25}: Kitamoto et al.\ 1992; Terada et al.\
1998); \astrobj{GS1124-68}{GS\,1124$-$68}: Kitamoto et al.\
1992; Ebisawa et al.\ 1994).
Moreover, the timescale between the primary and secondary maximum of the 
{\it GRO} BATSE hard X-ray outburst light curve of 
\astrobj{GROJ0422+32}{GRO\,J0422+32} as derived by the model of Augusteijn et
al.\ (1993) is similar to the time of appearance of the first `minioutburst'
with respect to the end of the X-ray outburst and the time between the two
`minioutbursts' (see Callanan et al.\ 1995, Chevalier \&\ Ilovaisky 1995).

\subsection{X-ray spectral hardening}

For the first time we have demonstrated that \astrobj{A0620-00}{A\,0620$-$00} 
exhibited considerable hardening $\sim$100~days after the start of the outburst.
A similar hardening at nearly the same time after outburst maximum 
has also been seen in \astrobj{GS1124-68}{GS\,1124$-$68} and 
\astrobj{GS2000+25}{GS\,2000+25} 
(Kitamoto et al.\ 1992; Ebisawa et al.\ 1994; 
Terada et al.\ 1998). The X-ray spectral and power spectral 
behaviour in \astrobj{GS1124-68}{GS\,1124$-$68} and 
\astrobj{GS2000+25}{GS\,2000+25} just before the hardening is consistent with 
canonical black-hole high-state behaviour (and maybe also for
\astrobj{A0620-00}{A\,0620$-$00}, see Section~\ref{a0620_intro}),
whereas after the hardening it is consistent with the canonical
black-hole low-state behaviour (Ebisawa et al.\ 1994; 
Miyamoto et al.\ 1994; Terada et al.\ 1998). 
We therefore suggest that the 
power-spectral behaviour after the start of the spectral hardening in 
\astrobj{A0620-00}{A\,0620$-$00} might have shown low-state like behaviour as 
well.

The time of maximum X-ray spectral hardening is close to when the 
$\sim$7.8-day modulations in the X-ray light curve of 
\astrobj{A0620-00}{A\,0620$-$00}
are strongest. Similarly, the time of maximum X-ray spectral 
hardening in \astrobj{GS1124-68}{GS\,1124$-$68} and 
\astrobj{GS2000+25}{GS\,2000+25} also occur simultaneously with 
the drops in intensity, i.e.\ $\sim$150~days after the start of the 
outburst. This suggests a connection between the X-ray spectral hardening 
and the occurrence of periodic modulations or drops in intensity
(see also Section~\ref{dips}).

Recently, a self-consistent model of accretion flows around black holes 
with various \.M has been put forward by Esin, McClintock \&\ Narayan 
(1997; see also Esin et al.\ 1998). Their accretion flow 
consists of an ADAF and an 
outer (thin) accretion disk. In addition above the orbital plane there 
is a hot corona. \.M determines the size of the ADAF region and the
density of the corona. In this way they could explain the different
states seen in black-hole binaries, the off-state, low state, high state
and (possibly) the very-high state. The spectral hardening $\sim$150
days after the outburst is identified with the transition from 
the high state to low state, i.e.\ the intermediate state in their 
model.

We note that such hardening is probably not confined to black-hole SXTs. 
It has recently been suggested that the hardening near the end of the outburst
of the neutron star SXT Aql\,X-1 (Zhang, Yu \&\ Zhang 1998) 
after a small secondary maximum is a similar phenomenon 
(Shahbaz et al.\ 1998a). If 
this is true, this should manifest itself in a change in the power 
spectral shapes, possibly resembling black-hole low-state behaviour 
during the hardening.

\subsection{Dips and intermediate maxima}\label{dips}

During the outburst the X-ray light curves show that various drops in 
intensity occurred, lasting from about one day to several days. 
They occurred $\sim$87, $\sim$103, and $\sim$228~days after the start 
of the outburst. Possibly, also the precursor may in fact represent a dip.
In addition to these dips, periodic drops in intensity have been 
seen in X-ray and optical with a period of about 7.8~days 
(see Section~\ref{a0620_intro}), and possibly also in the infra-red
(this paper). The occurrence of the dips are consistent with 
the $\sim$7.8~day modulation period.

It has been suggested that the $\sim$7.8-day modulation is the beat 
period between the orbital period and the (possible) superhump period
(Zhang \&\ Chen 1992), i.e.\ the disk precession period
(see e.g.\ Priedhorsky \&\ Holt 1987). 
In the SXTs where superhumps have been seen 
the period excess was found to be between $\sim$1--2\%\ 
(O'Donoghue \&\ Charles 1996). This has also been found for the 
TOADs (see Kuulkers, Howell \&\ van Paradijs 1996, and references therein), and 
has been related to the extreme mass ratio in these systems. 
If this also holds for \astrobj{A0620-00}{A\,0620$-$00}, the estimated 
superhump period would 
be 7.83--7.90\,hrs, which leads to a beat period between 17--32~days.
This is consistent with estimates of the beat period period from
its relation with the orbital period and the mass ratio (see e.g.\ Warner 1995),
i.e.\ $\sim$18--19~days. Hence the $\sim$7.8-day modulation cannot
be related to the beat period between the orbital period and the (possible)
superhump period.

Alternative models include e.g.\ intrinsic oscillations of the companion 
star, modulating the accreted matter onto the black hole 
(Ciatti et al.\ 1977).
We note that near the time of the strong $\sim$7.8-day modulations
the X-ray spectra harden (see previous Section), which suggest a connection
between the two. It might be that the transition radius between the outer thin disk
and inner ADAF flow in the model of Esin et al.\ (1997) is 
oscillating around that time, i.e.\ the transition radius moves inwards and
outwards on a time scale of $\sim$8~days, which therefore modulates
the accretion rate and subsequently the amount of X-rays.
Another speculation might be that the heating/cooling waves oscillate inwards
and outwards between certain radii, as a 
result of a modulation in the strength of X-ray irradiation. It remains to be 
seen if such mechanisms do indeed exist. Moreover, the delay in optical and 
X-rays should also be explained in such models.

The dip in X-rays $\sim$87~days after the start of the outburst has not 
been noted before. As we have shown this corresponds to a short 
{\it rise} in intensity or flare at hard energies ($\gtrsim$6\,keV),
which indicates the X-ray spectrum suddenly pivots for a short time.
In the optical we see that there may be a depression in the light curve, 
but shortly afterward the brightness increased above the expected
exponential decline, which resulted in another local maximum, 
a so-called `intermediate' maximum, $\sim$30~days after the secondary 
maximum. A similar `intermediate' maximum may be present in the infra-red, 
whereas there is {\it no} indication for such a maximum in X-rays.

The $\sim$30~days interval between the secondary maximum and `intermediate'
maximum is comparable to the viscous time scale in an irradiated 
accretion disk with parameters appropriate for A\,0620$-$00 
(King \&\ Ritter 1998; see also Shahbaz et al.\ 1998b). If irradiation triggers 
a thermal instability in the outer accretion disk, new material will 
accrete one viscous time scale later. This has been put forward as the 
explanation for the secondary maximum as an echo of the primary maximum 
by King \&\ Ritter (1998; we refer to e.g.\ Cannizzo 1998 and
references therein for other models on the cause of the secondary maximum). 
We suggest that the 
secondary maximum induces again enhanced irradiation which may then
show up one viscous time scale later. It is not clear to us, however, why 
this would cause the X-ray spectrum to pivot around that time, and why it 
does not show up as an `intermediate' maximum in the X-ray light curve.

The only other SXT where another local maximum has been reported which
did not coincide with a local maximum in X-rays is 
\astrobj{GROJ0422+32}{GRO\,J0422+32}.
Shrader et al.\ (1994) reported a secondary maximum in the ultra-violet,
which occurred $\sim$38~days after the maximum of the {\it GRO} BATSE outburst. 
However, the {\it GRO} BATSE light curve is that 
for high ($\gtrsim$20\,keV) energies. For GS\,1124$-$68 
the soft X-ray ($\sim$1--10\,keV) light curves differs from that in 
hard X-rays ($\sim$10--40\,keV light curves (see Ebisawa et al.\ 1994), and
this could also apply to \astrobj{GROJ0422+32}{GRO\,J0422+32}. 
The local ultra-violet maximum might therefore have been simultaneous with a 
local soft X-ray maximum, which is expected to exist $\sim$35~days after 
outburst maximum (Shahbaz et al.\ 1998b).

Although the time of the `hiccup' at the end of the outburst is consistent with 
the minima expected in the $\sim$7.8~days modulation, interestingly the 
time difference between the optical maximum of the tertiary peak and the
`hiccup' is also on the order of 30~days. This may suggest a
similar cause for the `hiccup' as that proposed for the `intermediate' 
maximum, where now the `hiccup' is a response to the tertiary maximum. 
We note that a similar `hiccup' might have been seen at the end of the outburst
of \astrobj{GS2000+25}{GS\,2000+25} (Chevalier \&\ Ilovaisky 1990).

\subsubsection{TOADs and SXTs}\label{TOADs}

We have shown here that the outburst light curve of the SXT 
\astrobj{A0620-00}{A\,0620$-$00} is very similar in shape to that of the 
TOAD \astrobj{ALCom}{AL\,Com}. Also the structure of the 1978 outburst light 
curve of \astrobj{WZSge}{WZ\,Sge} (Patterson et al.\ 1981) is very 
similar to that of \astrobj{ALCom}{AL\,Com} (see 
Howell et al.\ 1996) and \astrobj{A0620-00}{A\,0620$-$00}, having a similar 
exponential decay, drops in magnitude near the end of the outburst and 
a `bump' at the end of the outburst. 
The timescales in the outbursts of \astrobj{A0620-00}{A\,0620$-$00} and 
\astrobj{ALCom}{AL\,Com} differ by a factor of $\sim$5.3. This is close to the 
ratio of the orbital periods of \astrobj{A0620-00}{A\,0620$-$00} and 
\astrobj{ALCom}{AL\,Com} ($\sim$84~min, see Howell et al.\ 1996), 
i.e.\ $\sim$5.5. We note that the mass ratio of \astrobj{ALCom}{AL\,Com} is 
close to \astrobj{A0620-00}{A\,0620$-$00}, i.e.\ $q$$\lesssim$$0.15$, with 
likely values between 0.033--0.075 
(Howell, Hauschildt \&\ Dhillon 1998).
This shows that the optical outburst light curve is not governed by 
the mass of the compact object, but related to the (similar) 
disk properties.

Interestingly, the morphology of the orbital light curve of 
\astrobj{ALCom}{AL\,Com} in quiescence is rather unstable 
(Abbott et al.\ 1992; Howell et al.\ 
1996, and references therein), which has also been reported for 
\astrobj{A0620-00}{A\,0620$-$00} by Haswell (1996) and 
Leibowitz, Hemar \&\ Orio (1998). We note that the optical spectra 
of \astrobj{A0620-00}{A\,0620$-$00} in quiescence have also been reported to 
change with time (Murdin et al.\ 1980; Orosz et al.\ 1994).

Masetti \&\ Reg\H os (1997) suggested that SXTs 
share properties with another \astrobj{SUUMa}{SU\,UMa} subclass, i.e.\ 
the so-called \astrobj{ERUMa}{ER\,UMa} stars. This was based solely on the 
comparison of the shape and  appearance of (super) humps in the light curves 
of SXTs and \astrobj{ERUMa}{ER\,UMa} stars. \astrobj{ERUMa}{ER\,UMa} stars 
are at the other extreme of the \astrobj{SUUMa}{SU\,UMa} class, i.e.\
during quiescence they have highest mass transfer rates compared to
other \astrobj{SUUMa}{SU\,UMa} stars in quiescence. 
They therefore show very frequent outbursts with short quiescent periods,
40--50~days, while their outbursts are of relatively small amplitude,
$\Delta$V$\sim$3 (Kato \&\ Kunjaya (1995).
These are properties clearly not shared by the SXTs.

The outburst and quiescence properties of SXTs have been shown, however, 
to be very similar to those of TOADs, having a fast rise, large outburst 
amplitude, a slow outburst decay, drops in intensity near the end of the 
main outburst, and/or post-outburst brightenings
(Kuulkers et al.\ 1996). 
Our comparison between \astrobj{A0620-00}{A\,0620$-$00} and 
\astrobj{ALCom}{AL\,Com} gives additional
support for their similarity.
The similarities between the TOADs and SXTs reflect that both have 
low mass ratios and very low mass transfer rates, \.{M} 
(Kuulkers et al.\ 1996). 

It is clear that these light curves represent some very generic behaviour. 
This may be due to a pure viscous evolution, i.e.\ the outburst disk
just evolves under a hot-state viscosity, without the intervention of cooling 
fronts, until possibly the end of the outburst. 
Hence any mechanism that prevents the cooling front travelling in from the
outer edge of the disk for a sufficiently long time will produce these
light curves. In the SXTs irradiation naturally provides that mechanism
(e.g.\ Chen et al.\ 1993; Van Paradijs 1996; King 1998;
King \&\ Ritter 1998),
whose presence is suggested (at least near maximum of the outburst) by 
observations (see e.g.\ van Paradijs \&\ McClintock 1994; 1995;
Shahbaz \&\ Kuulkers 1998)\footnote{The amount of optical radiation reprocessed
from X-ray irradiation is still a subject of debate (see e.g.\ 
Lasota \&\ Hameury 1998). Current versions of the 
disk-instability model suggest that {\it only} $\sim$1~mag of the optical light
during the first months of the outburst is due to X-ray irradiation
(see Cannizzo 1998).}.
For the outburst of AL\,Com irradiation may prevent the 
cooling front from propagating as well, but it
is more complicated in this case (see King 1997). 
If confirmed by more detailed modelling, it may be possible to provide a 
unified explanation of both cases.

\section*{Acknowledgements}

First of all, we would like to thank Ron~Webbink for providing his 
compilation of optical observations of A\,0620$-$00.
Discussions with him and Rich Wolfson related to the first optical 
observations are also acknowledged.
We also acknowledge discussions with Phil Charles and Andrew King during 
the gestation of this paper, Bob Argyle for measuring the two additional 
archival plates taken with the Herstmonceux 26-inch telescope of the Royal 
Greenwich Observatory, and Roelf Takens for providing articles which were not 
available in the library of the Nuclear \&\ Astrophysics Laboratory.

In this research, we have used, and acknowledge with
thanks, data from the AAVSO International Database,
based on observations submitted to the AAVSO by variable
star observers worldwide. We also acknowledge the use of observations made 
by various variable star observers of the VSS~RAS~NZ, VSOLJ, and the AFOEV,
and thank Frank Bateson for providing the 
VSS~RAS~NZ measurements and Taichi Kato for providing the 
VSOLJ measurements.
We gratefully acknowledge the use of the processed {\it SAS-3} data 
from Kenneth Plaks, Jonathan Woo and George Clark.
We also thank Wan Chen for providing the {\it Ariel V} ASM and part of 
{\it Ariel V} SSE data points for A\,0620$-$00, as well as the 
X-ray data of the other SXTs, Nicola Masetti for providing part of the 
optical data of GRS\,1009$-$45, and Darragh O'Donoghue for most of the 
optical data of GS\,2000+25 and part of that of GS\,1124$-$68.

This research has made use of the Simbad database, operated at CDS, 
Strasbourg, France.

\end{document}